\documentclass[article,notitlepage,preprintnumbers,superscriptaddress,nofootinbib]{revtex4-1}
\usepackage[utf8]{inputenc}
\usepackage[a4paper, total={7in, 9in}]{geometry}
\usepackage{amsmath}
\usepackage{amssymb}
\usepackage[dvipsnames]{xcolor}

\usepackage{enumerate}
\usepackage{amsmath}
\usepackage{tikz}
\usepackage[compat=1.1.0]{tikz-feynman}
\usepackage{feynmf}
\usepackage{slashed}
\usepackage{braket}
\usepackage{url}
\usepackage{natbib}
\usepackage{graphicx}
\usepackage{comment}
\usepackage{multirow}
\usepackage{mathrsfs,amssymb}  
\usepackage{cancel}
\usepackage[normalem]{ulem}

\definecolor{Zpurple}{RGB}{119, 50, 168}
\definecolor{niceblue}{rgb}{0.388235, 0.627451, 0.847059}
\definecolor{nicered}{rgb}{0.7,0.1,0.1}
\definecolor{nicegreen}{rgb}{0.1,0.5,0.1}
\definecolor{persianblue}{rgb}{0.11, 0.22, 0.73}

\usepackage[pdftex,bookmarks,linktocpage,pdfpagelabels,plainpages=false,hyperfigures,linkcolor=blue,citecolor=blue]{hyperref} 
\hypersetup{colorlinks=true}

\usepackage{atbegshi,picture}
\AtBeginShipoutNext{\AtBeginShipoutUpperLeft{%
  \put(\dimexpr\paperwidth-1cm\relax,-1.5cm){\makebox[0pt][r]{\footnotesize \texttt{CETUP-2023-023}}}%
}}

\begin{document}
\bibliographystyle{apsrev4-1}

\title{Insights from Binary Pulsars and Laboratories into Baryon Number Violation: Implications for GeV Dark Matter}

\author{Rouzbeh Allahverdi}
\email{rouzbeh@unm.edu}
\affiliation{Department of Physics \& Astronomy, University of New Mexico,~Albuquerque,~NM~87106,~USA}

\author{Adrian Thompson}
\email{a.thompson@northwestern.edu}
\affiliation{Northwestern~University,~Evanston,~IL~60208,~USA}

\author{Mohammadreza Zakeri}
\email{m.zakeri@eku.edu}
\affiliation{Department of Physics and Astronomy, University of Kentucky,~Lexington,~KY~40506,~USA}
\affiliation{Department of Physics, Geosciences, and Astronomy, Eastern Kentucky University,~Richmond,~KY~40475,~USA}

\begin{abstract}
    Rare 
    processes in laboratory and within astrophysical environments can be highly sensitive probes of baryon-number violating interactions at the TeV scale.
    We demonstrate the power of neutron stars to constrain baryon number violation by considering a minimal extension of the standard model involving a TeV-mass scalar mediator and a GeV scale Majorana fermion $\psi$. We find that a $\Delta B = 2$ mass-loss process in binary pulsar systems via $n \to \gamma \psi$ and the subsequent scattering $\psi n \to \pi^- K^+$ places stringent constraints on the model parameter space. 
    These limits will become much stronger, due to the possibility of $\Lambda \rightarrow \gamma \psi$ decays at the tree level, if the neutron star equation of state is hyperonic.
    We compare these constraints with ongoing and future collider 
    experiments, $n-\bar{n}$ oscillations, and dinucleon decay searches 
    at future large-scale neutrino experiments, finding that the binary pulsars bounds on couplings are significantly tighter for specific flavor combinations.
\end{abstract}

\maketitle

\section{Introduction}

Explaining the matter-antimatter asymmetry in the universe is one of the important problems at the interface of particle physics and cosmology (for a review, see~\cite{Canetti:2012zc}). Baryon number violation (BNV) is one of the three required conditions 
for generating a baryon asymmetry in the early universe~\cite{Sakharov}. Models of TeV-scale baryogenesis
provide interesting opportunities to probe the physics 
responsible for the observed matter-antimatter asymmetry in the
laboratory~\cite{Babu:2006xc,Babu:2006wz}. The BNV interactions can lead to distinct collider signals such as monojets and/or monotops~\cite{DGK}, 
as well as $\Delta B = 2$ interactions among nuecleons like dinucleon decay $p p \rightarrow K^+ K^+$ and neutron-anti-neutron ($n - {\bar n}$) oscillations~\cite{Proceedings:2020nzz, Rao:1982gt}. 

Our focus here is on models of BNV that involve hadrophilic GeV-scale particles. These models provide particularly interesting opportunities to probe BNV and tightly constrain it by considering neutron stars. The important point is that short-range repulsive forces at very high densities elevate the ground state energy of nucleons in the inner regions of neutron starts above that in the vacuum. The ground state energy depends on the equation of state of neutron stars, which can be as high as $\sim 1.5$ in the case of neutrons. This leads to new channels for nucleon decay, including GeV scale particles, inside neutron stars that are kinematically blocked in the vacuum.  

As a specific example, we consider a minimal extension of the standard model (SM) that involves a GeV scale Majorana fermion $\psi$ with BNV couplings to quarks mediated by a TeV-mass scalar. One can then have BNV decays $n,\Lambda \to \gamma \psi$ accompanied by scattering $\psi n \to \pi^- K^+$ inside neutron stars. The effective $\Delta B = 2$ process amounts to conversion of mass into free-streaming neutrinos, which affects the orbital period of neutron star binaries. Using three well-studied binary systems, we will impose tight limits on the strength of BNV interactions~\cite{Berryman:2022zic, Berryman:2023rmh, Zakeri:2023xyj, Gardner:2023wyl}. For hyperonic equations of state, these bounds are a few orders of magnitude stronger than those from dinucleon decay for specific flavor combinations.                     

This paper is organized as follows. In \S~\ref{sec:theory} we introduce a simple model of BNV that connects 
dark matter (DM) and baryogenesis, and we match its effective interactions with the 
Standard Model (SM) to the chiral perturbation theory (ChPT).
Lagrangian.
In \S~\ref{sec:lab} we review the landscape of laboratory tests of this model and suggest a potential channel for direct detection of DM. Then, in \S~\ref{sec:binary_pulsars}, we show how the baryon-number-violating interactions of the model manifest in neutron stars and binary pulsar systems, working out specific constraints on the model parameter space. In \S~\ref{sec:discussion} we examine the combined set of limits from binary pulsars and laboratory searches to highlight the complementarity between the two efforts. Finally, in \S~\ref{sec:conclusion} we summarize the results and remark on the general impact of binary pulsars on other baryon-number-violating models and the outlook for future searches.

\section{Theoretical Framework}\label{sec:theory}

We investigate a simple model of BNV that was first proposed in Ref.~\cite{Allahverdi:2013mza}, and studied in more detail in~\cite{Allahverdi:2017edd}, to explain the observed baryon asymmetry and the DM content of the universe and also address the baryon-DM coincidence puzzle. It is similar in nature to several others studied in the literature~\cite{Dev:2015uca, Davoudiasl:2015jja, Allahverdi:2010im,Jin:2018moh,Heeck:2020nbq}.
The model is a minimal extension of the SM that introduces
two color-triplet, iso-singlet scalars $X_1$, $X_2$ 
and a Majorana fermion $\psi$. The scalars $X_{1,2}$ have hypercharge $+4/3$ and couple to the fermion $\psi$ and to the right-handed (RH) up quark generations (with hypercharge $+4/3$) through baryon-number violating interactions at $\Delta B = 1$ level. The $X_{1,2}$ couple to the RH down quarks $d_i$, $d_j$ of generations $i,j$ (each with hypercharge $-2/3$).
In the 4-component chiral/Weyl representation, the interaction Lagrangian reads
\begin{equation}\label{eq:lagr}
\begin{split}
\mathcal{L} &\supset 
\lambda_i \delta_{\alpha \beta} \left( X^{\alpha} \bar{u}_i^{\beta} P_L \psi
+ X^{*, \alpha} \bar{\psi} P_R u_i^{\beta} \right)
+ 
\lambda_{ij}^\prime \epsilon_{\alpha \beta \gamma} \left( X^{*, \alpha} \bar{d}^{\beta}_i P_L d_j^{c, \gamma} 
+ X^{\alpha} \bar{d}^{c, \beta}_j P_R d_i^{\gamma}\right)
\end{split}
\end{equation}
where $i,j \in \{1,2,3\}$ are generation indices for the up-type quarks $u$ and down-type quarks $d$, respectively, and Greek indices represent $SU(3)$ color. For simplicity, we have included one species of $X$ only. 
The charge conjugation operator is $C = - i\gamma^2 \gamma^0$, such that $d^c = C(\bar{d})^T$, and $\bar{d}^c = d^T C$. Given that we do not observe quarks with open color, summation over the color indices in the second term on the right hand side of Eq.~\eqref{eq:lagr} implies that only the off-diagonal elements of the coupling matrix $\lambda^{\prime}_{ij}$ are relevant.   

The model in Eq.~\eqref{eq:lagr}
can lead to successful baryogenesis from the interference between two species of $X$ in their two-body decays in the presence of
CP-violating phases in $\lambda_i$ and $\lambda^{\prime}_{ij}$ couplings.
Moreover, $\psi$ is stable (and hence the DM candidate) if $m_{\rm p} - m_{\rm e} \leq m_\psi \leq m_{\rm p} + m_{\rm e}$~\cite{Allahverdi:2013mza}\footnote{As we will 
show, the upper limit can be somewhat extended for sufficiently small values of $\lambda_i \lambda_{jk}^{\prime}$ for which the lifetime of $\psi$ satisfies observational limits on decaying DM.}. The fact that $m_\psi \approx m_{\rm p}$ can in addition help us address the baryon-DM coincidence puzzle by explaining a dark matter abundance $\Omega_\text{DM} \approx 5 \Omega_B$. 

The BNV interactions in this model lead to interesting collider signals as well as observable predictions for $\Delta B = 2$ processes at low energies. We 
discuss these experimental signatures and the existing constraints in detail in the next section. 

In the remainder of this paper, we focus on the 
BNV interactions in this model in the context of astrophysical environments, notably neutron stars. For this purpose, we simplify the model by considering only the lighter $X$ with $m_{X} \sim \mathcal{O}(1 \, \text{TeV})$, which is the current lower bound from the LHC.
Secondly, we consider the case where $m_\psi \sim {\cal O}({\rm GeV})$, 
though not necessarily in the aforementioned mass range,
so that $\psi$ is kinematically accessible inside neutron stars. 
Finally, for simplicity,
we neglect CP-violating effects in this analysis and treat 
$\lambda_i$ and $\lambda_{ij}^\prime$ as real-valued parameters. 

We first 
obtain the spectrum of operator interactions with the fermion $\psi$ and the baryons and mesons from the quark-level effective interaction $\psi u_k d_i d_j$. This operator is similar in form to other baryon-number violating models motivated by, e.g., the neutron lifetime anomaly or $B$-mesogenesis~\cite{Alonso-Alvarez:2021oaj, Alonso-Alvarez:2021qfd}, that utilized ChPT to investigate low energy physics. By matching to equivalent operators in the 
ChPT Lagrangian,
we can obtain operators that give rise to a myriad of channels like the process $\psi n \to \pi^- K^+$ and other scatterings akin to induced nucleon decay (see also Refs.~\cite{Davoudiasl:2015mcm, Huang:2013xfa, Berger:2023ccd}). We follow 
Refs.~\cite{Claudson:1981gh, Davoudiasl:2011fj, Alonso-Alvarez:2021oaj} to perform the operator matching. For example, consider the lowest generational coupling combination $\lambda_1$, $\lambda_{12}^\prime$ given in our Lagrangian in 
Eq.~\eqref{eq:lagr}.
Integrating out the heavy $X$ field yields the tree-level mixing as a 4-fermion interaction, in two-component notation;
\begin{equation}
\label{eq:dim6_quark_lag_tree}
    \mathcal{L}_6 \supset \dfrac{\lambda_1 \lambda^\prime_{12}}{m_X^2} [\psi^\dagger u_R] [d^T_R (i \sigma_2) s_R].
\end{equation}
From this operator we derive the operators at the level of baryons and mesons. Omitting color indices, matching the operator in Eq.~\eqref{eq:dim6_quark_lag_tree}
to the $SU(3)_V$ chiral representation gives a relationship
\begin{equation}
    O_{ij} \equiv \dfrac{1}{2}\epsilon_{jkl} [q_{R,k}^T (i\sigma_2) q_{R,l}][\psi^\dagger \, q_{R,i}] \Longleftrightarrow \mathcal{L}_6 = \text{Tr}[\hat{C}^R O],
\end{equation}
where $\hat{C}^R$ is the 3$\times$3 spurion matrix and the $SU(3)_V$ latin flavor indices correspond as $u=1, d=2, s=3$. The $u$-$d$-$s$ combination in 
Eq.~\eqref{eq:dim6_quark_lag_tree} corresponds to $O_{11}$;
\begin{equation}
    \hat{C}^R[(ds)u] = \dfrac{\lambda^\prime_{12} \lambda_1}{m_X^2}  \begin{pmatrix}
1 & 0 & 0\\
0 & 0 & 0\\
0 & 0 & 0
\end{pmatrix}
\end{equation}
The only $SU(3)_V$ invariant combination of our spurion matrix with the baryon and meson octets is
\begin{equation}
\label{eq:chiPT_lagr}
    \mathcal{L}_\text{eff,ChPT}^{(0)} = \beta \text{Tr} [ \hat{C}^R u^\dagger B_R \psi u]\, ,
\end{equation}
where the spinor combination $B_R \psi$ is a shorthand for $b_{R}^{\dagger}\left[-i\sigma^2\right]\psi_{R}^* = \bar{b} P_L \psi^c$ and $u = e^{i \Phi / f_\pi}$. $\Phi$ and $B_R$ are the meson and baryon octets, respectively, defined below.
\begin{equation}
    \Phi =
    \begin{pmatrix}
        \frac{\pi^0}{\sqrt{2}} + \frac{\eta_8}{\sqrt{6}} & \pi^+ & K^+\\
        \pi^- & -\frac{\pi^0}{\sqrt{2}} + \frac{\eta_8}{\sqrt{6}} & K^0\\
        K^- & \overline{K}^0 & -\frac{2\eta_8}{\sqrt{6}}
    \end{pmatrix}, \, \, \, 
    B =
    \begin{pmatrix}
        \frac{\Lambda^0}{\sqrt{6}} + \frac{\Sigma^0}{\sqrt{2}} & \Sigma^+ & p\\
        \Sigma^- & \frac{\Lambda^0}{\sqrt{6}} - \frac{\Sigma^0}{\sqrt{2}} & n\\
        \Xi^- & \Xi^0 & -\frac{2\Lambda^0}{\sqrt{6}}
    \end{pmatrix},
\end{equation}
The factor $\beta$ in 
Eq.~\eqref{eq:chiPT_lagr} is computed in lattice QCD and taken as $\beta = 0.014$ GeV$^3$~\cite{JLQCD:1999dld}. By expanding 
Eq.~\eqref{eq:chiPT_lagr}
in powers of $1/f_\pi$, we obtain a set of interaction operators between $\psi$ and the SM baryons and mesons, which are shown in Appendix~\ref{app:chiral:expansion}. In particular, we recover the mixing term shown in 
Eq.~\eqref{eq:dense:lag_mix} by expanding 
Eq.~\eqref{eq:chiPT_lagr} to zeroth-order in $1/f_\pi$;
\begin{equation}
    \mathcal{L}_\text{eff,ChPT}^{(0)} \supset  \beta\dfrac{ \lambda_{1}\lambda^\prime_{12}}{m_X^2} \bigg( \frac{1}{\sqrt{6}}\bar{\psi}^c P_R \Lambda\bigg)
\end{equation}
where we would interpret $\varepsilon_{\Lambda\psi}^\textrm{tree} = \beta\dfrac{\lambda_{1}\lambda^\prime_{12}}{\sqrt{6} m_X^2}$. For the loop-level mixings to $\Lambda$ shown in Fig.~\ref{fig:n:mix}, we obtain the same spurion matrix texture but with the corresponding loop factors in 
Eq.~\eqref{eq:n_mixing}.
For the loop-level mixings to neutrons, the operator matching instead requires $O_{32}$.

\section{Laboratory Signals and Constraints}\label{sec:lab}
Given the model and framework in \S~\ref{sec:theory} we first discuss the existing limits from laboratory searches for the fermion $\psi$ and its BNV interactions, both through direct and indirect methods, and future improvement on these bounds. 

\subsection{Collider Searches}
Collider probes include searches for monojet/monotop signatures
\cite{Andrea:2011ws,Agram:2013wda,Alvarez:2013jqa,Wang:2011uxa,Kumar:2013jgb, Blanke:2020bsf}, which 
have the additional capability to distinguish left-chiral from right-chiral couplings~\cite{Allahverdi:2015mha}. Several searches have been performed at the Tevatron and the LHC with center-of-mass energies $\sqrt{s}=1.96$ TeV~\cite{CDF:2012obh}, 8 TeV~\cite{Khachatryan:2014uma,Aad:2014wza} and $13$ TeV~\cite{Sirunyan:2018gka}. 
Heavy scalar mediators are excluded for masses below 3.4 TeV in $\bar{d}_i\bar{d}_j \rightarrow \phi \rightarrow t \psi$ in events with large missing transverse momentum and a hadronically decaying top quark~\cite{Sirunyan:2018gka}. 
This leads to bounds excluding values of the coupling product on the order 
of $|\lambda_3 \lambda_{12}^\prime| \gtrsim 10^{-2}$.

A monojet search at CMS~\cite{Undleeb:2017oor} found limits on $\lambda_1$ and $\lambda_{12}^\prime$ for $m_\psi = 1$ GeV and $m_X = 1$ TeV. 
Interpreting this limit as a bound on the coupling product $| \lambda_{12}^\prime \lambda_1|$ leads to a range of bounds, since the sensitivity scaling with respect to these couplings does not vary trivially. However, one finds that this search was sensitive to coupling products in the range of $1.17 \times 10^{-1} \leq |\lambda_{12}^\prime \lambda_1| \leq 3 \times 10^{-3}$. This limit can only be expressed as a range of coupling products, dependent on the individual choices of $\lambda_{12}^\prime$ and $\lambda_1$.
Since the event rate is sensitive to the finite width of the mediator $X$ that drives the final state, we 
observe that the limit of large widths introduces a non-trivial dependence on $\lambda_{12}^\prime$ and $\lambda_1$ in the propagator. 
Additionally, searches that would be sensitive to higher generation couplings such as $\lambda_3 \lambda_{13}^\prime$ have been proposed~\cite{Kalsi:2024hfe}.

\subsection{Searches for Di-nucleon Decay}
Laboratory searches for BNV have also included investigations of the di-nucleon decay process $p p \to K^+ K^+$, which is enabled by the Majorana mass of $\psi$.
A search for this process was conducted at Super-Kamiokande~\cite{sk_kaons} using the water Cherenkov detector, passively inside $^{16}$O nuclei over a collected exposure of $3.1 \times 10^{33}$ oxygen-years.
From the bounds on the lifetime $\tau(pp \to K^+ K^+)$, we can derive a constraint on the couplings and masses of
particles in our model.

The decay width of $pp \to K^+ K^+$ has been previously estimated  in the literature, using simple dimensional analysis arguments, for models similar to the one we consider in this work~\cite{Goity:1994dq, Litos:2010zra}.
For instance, the operators proportional to $\lambda_1$ and $\lambda_{12}^\prime$ yield the result\footnote{These operators also lead to $\Delta S = 2$ processes of $K^0_s-K^0_{\bar s}$ and $B^0_s-B^0_{\bar s}$ mixing. However, color conservation does not allow any tree-level contribution to the mixing~\cite{BM}, and one-loop contributions easily satisfy the experimental constraints for the values of $\lambda \lambda^{\prime}$ that are allowed by di-nucleon decay bounds~\cite{Allahverdi:2010im}.}
\begin{equation}
    \Gamma(pp \to K^+ K^+) = \frac{8}{\pi}(\lambda_1 \lambda_{12}^\prime)^4 \frac{ \Lambda_\text{QCD}^{10} \rho_N}{m_p^2 m_\psi^2 m_X^8} ,
\end{equation}
where we take $\Lambda_\text{QCD} = 150$ MeV as the approximate QCD scale parameter, $\rho_N = 0.25$ fm$^{-3}$ is the nuclear density, and $m_p$ is the proton mass. The Super-K limit on the lifetime is $\tau / BR_{pp \to K^+ K^+}  < 1.7 \times 10^{32}$ years at 90\% confidence limit (C.L.) corresponding to 91.6 kton-years.
This excludes the coupling combination $\lambda_1 \lambda_{12}^\prime$ for a given $m_X$ mass with the relation
\begin{align}
\label{eq:sk_diproton}
    |\lambda_1 \lambda_{12}^\prime | &> 7.0 \times 10^{-8} \bigg(\frac{m_X}{\text{TeV}}\bigg)^2 \bigg(\frac{m_\psi}{\text{GeV}}\bigg)^{1/2}\, \, \, \, \, \, \, \, \, \,  \text{Super-K exclusion, 90\% C.L.}
\end{align}

The DUNE Far Detector (FD) complex~\cite{DUNE:2020ypp, DUNE:2020lwj} may also be sensitive to the same process, $p p \to K^+ K^+$, occurring in $^{40}$Ar nuclei. Its sensitivity can be estimated by resealing the Super-K limit, accounting for two key differences: the exposure and the nucleus of interest.
First, DUNE-FD is expected to have four detector modules,
each containing 10 kton of liquid argon.
Since not all modules may be completed 
simultaneously, we assume a conservative scenario in which two modules are active for a full 10 years and the 
remaining two for 6 years, resulting in a total exposure of 320 kton-years. Second, 
because the underlying physics 
driving $p p \to K^+ K^+$ is short-ranged, 
the decay width for this process should 
scale with the number of protons in the nucleus, $Z$.
Therefore, the rates in argon should be $\sim 18 / 8 = 2.25$ times larger than in $^{16}$O. Given that DUNE-FD is a liquid argon time projection chamber (LArTPC) with particle identification and tracking 
capabilities, we estimate its efficiency for this unique final state to be 100\%, or 5 times larger than
that of 
Super-K. Altogether, this scaling yields a 90\% C.L. sensitivity to
\begin{align}
\label{eq:dune_diproton}
    |\lambda_1 \lambda_{12}^\prime | &> 3.8 \times 10^{-8} \bigg(\frac{m_X}{\text{TeV}}\bigg)^2 \bigg(\frac{m_\psi}{\text{GeV}}\bigg)^{1/2}\, \, \, \, \, \, \, \, \, \,  \text{DUNE-FD sensitivity, 90\% C.L.}
\end{align}
Similarly, we can also perform a projection for Hyper-Kamiokande (Hyper-K), a future large-volume water Cherenkov detector~\cite{Hyper-Kamiokande:2016srs, Hyper-Kamiokande:2018ofw}. For a simple estimate, we assume a 10-year exposure with the same signal efficiency as
Super-K, but with a fiducial detector volume of 190 kton H$_2$O~\cite{AliAjmi:2024xus}. By rescaling the Super-K limit, we find a future sensitivity 
for Hyper-K of
\begin{align}
\label{eq:hk_diproton}
    |\lambda_1 \lambda_{12}^\prime | &> 2.35 \times 10^{-8} \bigg(\frac{m_X}{\text{TeV}}\bigg)^2 \bigg(\frac{m_\psi}{\text{GeV}}\bigg)^{1/2}\, \, \, \, \, \, \, \, \, \,  \text{Hyper-K exclusion, 90\% C.L.}
\end{align}
The exclusions from Super-K and projected sensitivities for Hyper-K and DUNE-FD in Eqs.~(\ref{eq:sk_diproton}, \ref{eq:dune_diproton}, \ref{eq:hk_diproton}) are all within an order of magnitude in coupling for fixed $m_X$ and $m_\psi$, despite approximately an order of magnitude of improved signal yield sensitivity at DUNE-FD and Hyper-K. This is because the decay rate scales with the fourth power of the coupling product. To see an order of magnitude improvement in the sensitivity to the coupling product, much larger decay volumes are needed.

\subsection{Neutron-antineutron Oscillations}
\label{sec:nnbar}
As described in Ref.~\cite{Allahverdi:2017edd}, the operators in Eq.~\eqref{eq:lagr}
can give rise to $n-\bar{n}$ oscillations. This can take place via the dimension-9 effective operator
\begin{equation}
    \mathcal{L}_{n\bar{n}} = \dfrac{(\lambda^\prime_{1j})^2 \lambda_1}{m_X^2} [\psi^\dagger u_R] [d^T_R (i \sigma_2) d^j_R] [d^T_R (i \sigma_2) d^j_R]^\dagger,
\end{equation}
where $j=2,3$ so $d_R^j = s_R, b_R$. The decay width for $n\to \bar{n}$ is approximately given by
\begin{equation}
    \Gamma_{n\bar{n}} \simeq \Lambda_\text{QCD}^6 \dfrac{\lambda_1^2 (\lambda_{1j}^\prime)^4 m_\psi}{16\pi^2 m_X^6} \ln \bigg(\frac{m_X^2}{m_\psi^2}\bigg),
\end{equation}
which we then use to constrain $\lambda_1$, and $\lambda_{12}^\prime$ or $\lambda_{13}^\prime$ for 
given $m_\psi$ and $m_X$ using the bounds on the $n-\bar{n}$ lifetime. SNO~\cite{SNO:2017pha} obtained a lower bound for the free oscillation lifetime $\tau_{n\bar{n}} > 1.23 \times 10^8$ s in bounded nuclei and $\tau_{n\bar{n}} > 1.37 \times 10^8$ s from an unbounded beam of cold neutrons. Additionally, they recast the Super-K result~\cite{PhysRevD.91.072006} for bound neutron oscillations to be $\tau_{n\bar{n}} > 2.9 \times 10^8$ s. More recently, a bound of $\tau_{n\bar{n}} > 4.7 \times 10^8$ s was obtained by Super-K~\cite{PhysRevD.103.012008}. All of these limits are taken at 90\% confidence limit. For this work, we 
take the more recent and conservative bound of $\tau_{n\bar{n}} > 4.7 \times 10^8$ s for the neutron oscillation lifetime and use it to constrain the coupling product $|\lambda_1 \lambda_{13}^\prime|$ for $m_X = 1$ TeV as a function of $m_\psi$. To do this, we scan over the couplings for several $m_\psi$ and search for the combination of $|\lambda_1 \lambda_{13}^\prime|_\text{max}$ such that all coupling products greater than this value would necessarily 
result in a larger lifetime than the bound (since the lifetime depends on the non-trivial combination $\lambda_1^2 (\lambda_{13}^\prime)^4$). From the scan, for all masses $m_p - m_e < m_\psi < 1.5$ GeV, we find a conservative exclusion:
\begin{align}
    |\lambda_1 \lambda_{13}^\prime | & \gtrsim 1.83 \times 10^{-3} \, \, \, \, \, \, \, \, \, \,  n-\bar{n}\text{ oscillation lifetime exclusion, 90\% C.L.}
\end{align}


\subsection{Dark Matter Direct Detection}
\begin{figure}[h]
    \centering
    \includegraphics[width=0.9\textwidth]{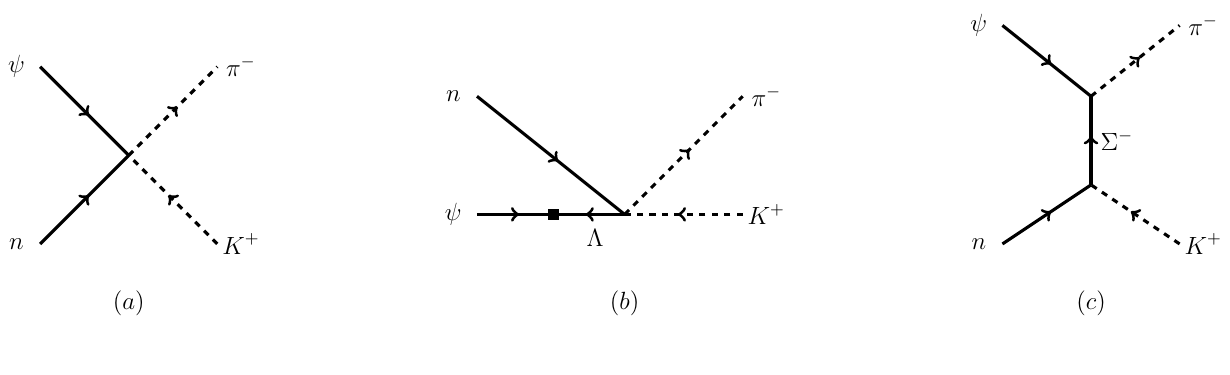}
    \caption{$\psi n \to K^+ \pi^-$ scattering amplitude contributions at $1/f^2_\pi$ order.}
    \label{fig:charged_meson_scattering_feynman}
\end{figure}

\begin{figure}
    \centering
    \includegraphics[width=0.6\textwidth]{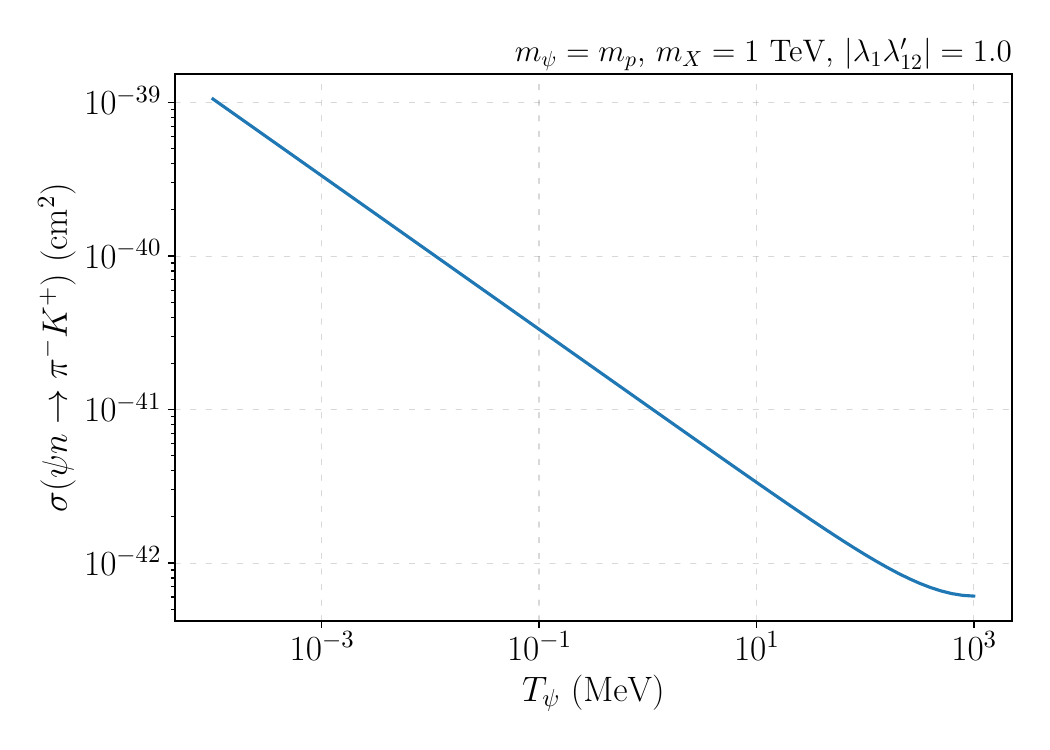}
    \caption{The inclusive scattering cross section as a function of the kinetic energy of $\psi$ for $\psi + n \to \pi^- + K^+$, calculated using the amplitudes shown in Fig.~\ref{fig:charged_meson_scattering_feynman}.}
    \label{fig:xs_neutron_psi}
\end{figure}

It was shown in \S~\ref{sec:theory} that expanding out 
Eq.~\eqref{eq:chiPT_lagr}
in powers of $1/f_\pi$, we obtain interaction terms between $\psi$ and the SM baryons and mesons. Together with the SM baryon-meson ChPT 
Lagrangian,
\begin{equation}
\begin{split}
    \mathcal{L}_{\phi B}^{(1)} &= \frac{D}{2} {\rm Tr} \left( \overline{B} \gamma^{\mu} \gamma_{5} \{u_{\mu}, B\}\right) +  \frac{F}{2} {\rm Tr} \left( \overline{B} \gamma^{\mu} \gamma_{5} [u_{\mu}, B]\right) \\
    &\supset \dfrac{(D+3F)}{f_\pi^2} \bigg[ \frac{i}{4\sqrt{3}} \bigg( (\overline{K^0} \partial_\mu \pi^0) \overline{\Lambda^0}\gamma^\mu \gamma^5 n  + (\pi^0 \partial_\mu K^0) \overline{n}\gamma^\mu \gamma^5 \Lambda^0 \bigg) \\
    & - \frac{i}{2\sqrt{6}} \bigg( (\pi^- \partial_\mu K^+) \overline{n}\gamma^\mu \gamma^5 \Lambda^0 + (K^-\partial_\mu \pi^+) \overline{\Lambda^0}\gamma^\mu \gamma^5 n \bigg)\bigg] \label{eq:chpt:Lagr}
\end{split}
\end{equation}
they give rise to operators relevant for $\psi n \to \pi^- K^+$ scattering, shown in Fig.~\ref{fig:charged_meson_scattering_feynman}. We take $D = 0.80$ and $F = 0.46$~\cite{Ledwig:2014rfa}. These are similar to other forms of induced nucleon decay (IND) that arise from similar BNV interactions~\cite{Davoudiasl:2011fj,Jin:2018moh}. We discuss these scattering processes further along with the computation of their cross sections in Appendix~\ref{app:scattering}. The total inclusive cross section for unit couplings is shown in Fig.~\ref{fig:xs_neutron_psi}. 
Additionally, similar processes involving scattering off protons may also occur.
From the expansion at order $1/f_\pi^2$, contact interactions allow for $\bar{\psi}^c p \to K^+ \pi^0$, $\bar{\psi}^c p \to K^+ \eta^8$,  and $\bar{\psi}^c p \to \bar{K}^0 \pi^+$, among others via 
hyperon mixing and 3-point vertices. The middle diagram in Fig.~\ref{fig:charged_meson_scattering_feynman} dominates this process 
due to the hyperon-$\psi$ mixing indicated by the black square. 

If $\psi$ 
constitutes the total relic abundance in the galactic halo, 
the scattering channels depicted in Fig.~\ref{fig:charged_meson_scattering_feynman} can be utilized to search for dark matter in large detector volumes like DUNE-FD. 
Recall that this scenario requires the stability of $\psi$ and the proton, constraining the mass 
to the range $m_p - m_e < m_\psi < m_p + m_e$. We could then look for the scattering of $\psi$ DM in the galactic halo inside a large detector via $\psi n \to K^+ \pi^-$, which would lead to a very unique detection signature. Specifically, this would lead to final states such as
$\psi n \to K^+(\to \mu^+ \nu_\mu) \pi^-(\to \mu^- \bar{\nu}_\mu)$ and $\psi n \to K^+(\to e^+ \nu_e) \pi^-(\to \mu^- \bar{\nu}_\mu)$.
This is particularly distinctive in a LAr-TPC type detector where the pion and kaon tracks could be reconstructed.
We forecast sensitivity of the DUNE FD to this type of DM scattering over the same exposure used above for the dinucleon decay search, with the local DM distribution function taken to be
\begin{equation}
    f_\psi(\vec{v}) = \dfrac{1}{N_\text{esc} \pi^{3/2} v_0^3} \exp \bigg( -\frac{(\vec{v}+\vec{v}_\oplus)^2}{v_0^2} \bigg) \Theta(v_\text{esc} - |\vec{v} + \vec{v}_\oplus|)
\end{equation}
where $N_\text{esc} = \text{erf} (v_\text{esc}/v_0) - \frac{2}{\sqrt{\pi}}(v_\text{esc}/v_0) \exp(-v_\text{esc}^2/v_0^2)$, $v_0 = 220$ km/s, $v_\text{esc} = 544$ km/s, and $v_\oplus = 244$ km/s~\cite{Smith:2006ym,Kerr:1986hz}. We 
expand the Gaussian exponent $(\vec{v}+\vec{v}_\oplus)^2$ as $v^2 + v_\oplus^2 + 2 v v_\oplus \cos\theta_\oplus$, where $\theta$ is the angle between the DM velocity $v$ and the speed of Earth in the galactic rest frame $v_\oplus$, which we take to be fixed. We then integrate to find the event rate:
\begin{equation}
\label{eq:dm_rate}
    R = T \bigg( \frac{\rho_\psi}{m_\psi} \bigg) \, 2\pi c\,  \int_0^\infty \int_{-1}^1 v^2  f_\psi(v) \, \sigma(v) \, d(\cos\theta_\oplus) dv 
\end{equation}
In 
Eq.~\eqref{eq:dm_rate},
$T$ is the time exposure in seconds, $c$ is the speed of light in cm/s, and $\sigma$ is the $\psi n \to K^+ \pi^-$ cross section on atomic $^{40}$Ar in cm$^2$, and $\rho_\psi$ is the DM energy density which we take to be $\rho_\psi = 0.4$ GeV/cm$^3$~\cite{Catena:2009mf}. 

Assuming a background-free search given the unique final state of this scattering process, we can take the background-free Poisson 90\% C.L. of 2.3 events to determine the sensitvity. Using the rate in 
Eq.~\eqref{eq:dm_rate}
integrated over an exposure of 320 kton years, we find a projected sensitivity on the coupling combination $|\lambda_1 \lambda_{12}^\prime|$ of
\begin{align}
    |\lambda_1 \lambda_{12}^\prime | &> 3.41 \times 10^{-7} \bigg(\frac{m_X}{\text{TeV}}\bigg)^2 \, \, \, \, \, \, \, \, \, \,  \text{DUNE-FD sensitivity to DM, 90\% C.L.}
\end{align}
This limit would not be as powerful as the limits from $p p \to K^+ K^+$ searches. However, a second possible manifestation of the model in 
Eq.~\eqref{eq:lagr}
is to let $\Psi$ be Dirac in nature and endowed with baryon number $B=1$. In this case, $X$ would also carry baryon number $B=-2/3$. In this scenario, the di-proton decay channel is no longer available, making the direct detection process $\psi n \to \pi^- K^+$ the most sensitive laboratory probe of the dark matter scenario.

\section{Effects on Binary Pulsars}\label{sec:binary_pulsars}
In this section, we explore the potential signatures of this model in neutron stars, focusing on the impact of baryon dark decays, such as ${\cal B} \to \psi \gamma$, on the orbital period of binary pulsars.
It is important to note that the baryon number and mass loss observed in the orbital dynamics of these binary systems is distinct from other baryon-number-violating signatures, such as the heating effects recently explored in Refs.~\cite{Ema:2024wqr, Fox:2024kda}.

The extreme conditions within neutron stars, where gravitational forces compress matter to densities exceeding nuclear saturation, offer a unique environment. At these densities, the repulsive core of nuclear interactions becomes significant, raising baryon ground-state energies and allowing for decay processes that are forbidden in a vacuum or in ordinary nuclei~\cite{Berryman:2023rmh}. Consequently, neutron stars act as natural laboratories for observing baryon dark decays~\cite{Gardner:2023wyl}, bypassing the  kinematic limitations in terrestrial experiments such as Super-Kamiokande and KamLAND~\cite{Super-Kamiokande:2015pys, KamLAND:2005pen}. This kinematic advantage is reflected in our exclusion curves in the $(m_{\psi}, \lambda_i \lambda_{jk})$ plane, which extend into regions where $m_{\psi} > m_{N}$. The effect is even more pronounced in heavier pulsars due to their higher core densities.

Given these kinematic advantages, the most massive pulsars, particularly those in binary systems, are prime candidates for observations.
However, in practice, the most massive pulsars are not always the most precisely timed, whereas the less massive but more precisely timed pulsars often provide tighter constraints due to the accuracy of their timing measurements. Moreover, the binary systems under consideration should have well-determined orbital parameters, small error bars on the mass, and should not be black-widow pulsars, where active mass transfer between the components could complicate the analysis. In our analysis, we focus on three candidate systems, detailed in Table~\ref{tab:psrbinary}.

Studying binary pulsars offers a distinct advantage: while BNV can influence pulsar spin characteristics—affecting spin-down rates, the second derivative of spin frequency, and braking indices—these effects are closely tied to the pulsar's magnetic field, which is typically inferred from the observed spin-down parameters~\cite{Zakeri:2023xyj}. In contrast, the orbital period changes in binary pulsars due to gravitational radiation are precisely determined and independent of the complex magnetic field modeling of the pulsar~\cite{Peters-GR, Stairs:2003eg, Hulse-Taylor-1975, Goldman:2009th, Goldman:2019dbq}. 
As measurement precision improves, the constraints on BNV effects in binary systems will become increasingly stringent, thereby enhancing our ability to utilize these systems as laboratories for testing fundamental physics~\cite{Gardner:2023wyl}.

The BNV processes we consider here are significantly slower than the typical Urca processes within neutron stars, leading to a quasi-equilibrium state where changes occur gradually. 
As demonstrated in Appendix~\ref{app:escapee}, the $\psi$ particles produced by baryons are too heavy to escape the neutron star's gravitational pull, rendering them gravitationally bound.  Consequently, their subsequent scattering off baryons, as depicted in Fig.~\ref{fig:charged_meson_scattering_feynman}, leads to additional baryon loss.
Given that the scattering channels can outpace the production rate of $\psi$, the resultant number density of $\psi$ within the system remains negligible. Consequently, this aligns with the ``depletion dominance'' framework outlined in Ref.~\cite{Gardner:2023wyl}. This framework posits a quasi-equilibrium evolutionary path for neutron stars, along a trajectory governed by a baryon-conserving EoS, as detailed in Ref.~\cite{Berryman:2022zic}. The impact of quasi-equilibrium BNV on the orbital period is briefly discussed in Appendix~\ref{app:formalism}. For a more comprehensive analysis, we direct the reader to Ref.~\cite{Gardner:2023wyl} and the references therein.

\begin{figure}[h]
 \centering
    \includegraphics[width=0.9\linewidth]{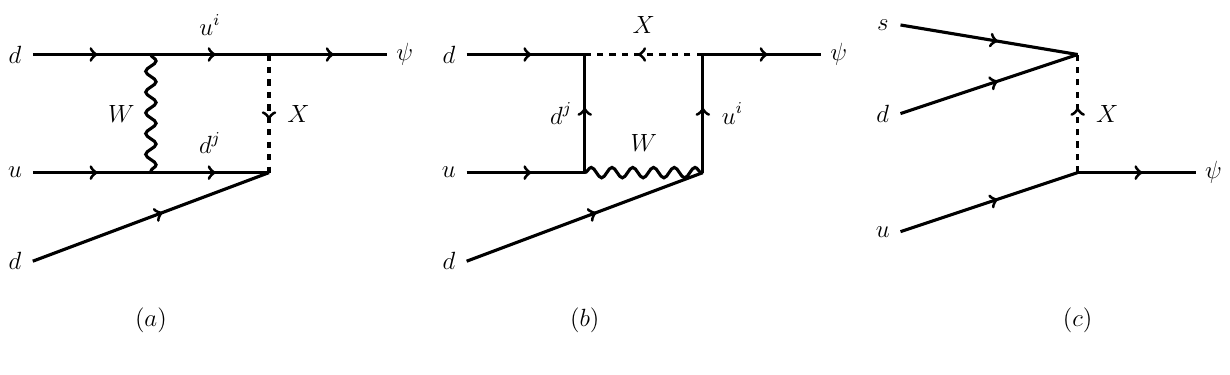}
    \caption{Illustration of the neutron-$\psi$ (a, b) and $\Lambda$-$\psi$ (c) mixing induced by interactions defined in the Lagrangian of Eq.~\eqref{eq:lagr}, and represented at low energies by Eq.~\eqref{eq:dense:lag_mix}. Note that we get additional diagrams for $\Lambda$-$\psi$ mixing if we replace the incoming down quark with a strange quark, $d \rightarrow s$, in diagrams (a) and (b). }
    \label{fig:n:mix}
\end{figure}

We employ the hadronic relativistic mean field theory, which provides a comprehensive framework for studying interactions under extreme densities~\cite{Walecka:1974qa,Serot:1984ey,Serot:1997xg,Dexheimer:2008ax}. The Lagrangian in Eq.~\eqref{eq:lagr} results in a mixing between neutral baryons and $\psi$, as illustrated in Fig.~\ref{fig:n:mix}. This interaction enables a set of baryon-number-violating processes within neutron stars' cores. We begin by examining the decay of neutrons and $\Lambda$ baryons through the process ${\cal B} \to \psi + \gamma$. The in-medium Lagrangian, which governs these interactions, is expressed by
\begin{equation}
\mathcal{L} = \overline{\cal B} \left( i
\slashed{\partial}
- \slashed{\Sigma}_{\cal B} - m_{\cal B}^* + \frac{g_{\cal B} e}{8m_{\cal B}^*} \sigma^{\alpha \beta} F_{\alpha \beta} \right) {\cal B} + \overline{\psi} \left( i
\slashed{\partial}
- m_{\psi} \right) \psi - \varepsilon_{{\cal B} \psi} \left( \overline{\cal B} P_L \psi + \overline{\psi} P_R \cal B\right) \,.\label{eq:dense:lag_mix}
\end{equation}
In this equation, \(m_{\cal B}^*\) represents the effective mass of the baryon, and \(\Sigma_{\cal B}^{\mu}\) denotes the baryon's self-energy. The parameter \(\varepsilon_{{\cal B} \psi}\) quantifies the mixing strength between the baryon and \(\psi\). The baryon's \(g\)-factor, \(g_{\cal B}\), is given as \(g_n = -3.826\) for neutrons and \(g_\Lambda = -1.22\) for \(\Lambda\) baryons~\cite{ParticleDataGroup:2022pth}. In this context, \(\sigma^{\alpha \beta} \equiv (i/2) [\gamma^{\alpha}, \gamma^{\beta}]\) is the commutator of the Dirac gamma matrices, \(F_{\alpha \beta}\) is the electromagnetic field strength tensor, and \(\slashed{\partial}\) represents the contraction of the Dirac gamma matrices \(\gamma_{\mu}\) with the spacetime derivatives \(\partial^{\mu}\).

The low-energy mixing parameter $\varepsilon_{{\cal B} \psi}$ can be expressed in terms of the couplings in Eq.~\eqref{eq:lagr}. Specifically, for neutrons, the mixing parameter is given by:
\begin{align}
\label{eq:n_mixing}
    \varepsilon_{n \psi} = 
    &\frac{G_F\, \gamma^L_{n}}{\sqrt{2}\, 4\pi^2\, m_W^2}
    \sum_{i} \sum_{j\neq 1} \lambda_{i} \lambda_{1j}' V_{i 1} V^{*}_{1 j} m_{d_j} m_{u_i}\,  F(x_{d_j}, x_{u_i}, x_{X}) \, ,
\end{align}
Similarly, for $\Lambda$-baryons, the loop-level contribution to the mixing parameter can be written as:
\begin{align}
\label{eq:lambda_mixing:loop}
    \varepsilon_{\Lambda \psi} = 
    &\frac{G_F\, \gamma^L_{\Lambda}}{\sqrt{2}\, 8\pi^2\, m_W^2}
    \sum_{i=1}^3 \lambda_{i}  m_{u_i} \left[ \sum_{j= 2}^{3}  \lambda_{1j}' V_{i 2} + \sum_{j= 1, 3} \lambda_{2j}' V_{i 1} \right] V^{*}_{1 j} m_{d_j}\,  F(x_{d_j}, x_{u_i}, x_{X}) \, ,
\end{align}
In addition to the loop-level contribution, the tree-level diagram for $\Lambda$-baryon gives the following expression for the mixing parameter:

\begin{align}
\label{eq:lambda_mixing:tree}
    \varepsilon_{\Lambda \psi} = 
    &\lambda_1 \lambda'_{12} a_{ds u}^{\Lambda} \frac{\beta}{m_X^2}.
\end{align}

We have defined $a_{ds u}^{\Lambda} = 1/\sqrt{6} $, 
$\gamma^L_n P_R u_n \equiv \langle 0 |\left[\epsilon_{\alpha\beta\gamma}\,  u^{\alpha T} C P_L d^{\beta}\right] \left[ P_R d^{\gamma}\right]  | n \rangle$, with $\gamma^L_n = -a_{n} \beta$, $a_n = 1$~\cite{Alonso-Alvarez:2021oaj}, and $\beta \approx 0.014(2)\, {\rm GeV}^3$~\cite{Aoki:2017puj}. The loop function $F(x_D, x_U, x_X)$ is defined by\footnote{We corrected that errors in the loop function $I(x_1, x_2, x_{\bar{S}_1})$ reported in equation (19) of Ref.~\cite{PhysRevD.103.055012}.}
\begin{equation}
\begin{split}
    F(x_D, x_U, x_X)  =&
    \frac{\left(4-x_D\right) x_D \ln \left(x_D\right)}{\left(1-x_D\right) \left(x_D-x_U\right) \left(x_D-x_X\right)}
   +
   \frac{\left(4-x_U\right) x_U\ln \left(x_U\right)}{\left(1-x_U\right) \left(x_U-x_D\right)\left(x_U-x_X\right)} \\
   &+
   \frac{\left(4-x_X\right) x_X \ln
   \left(x_X\right)}{\left(1-x_X\right) \left(x_X-x_D\right)
   \left(x_X-x_U\right)}, 
\end{split}
\end{equation}
where $x_D \equiv m_{d_j}^2 / m_W^2$, $x_U \equiv m_{u_i}^2 / m_W^2$, $x_X \equiv m_{X}^2 / m_W^2$.  

Having established the relationship between the mixing parameter $\varepsilon_{B\psi}$ and the model parameters $\lambda_i \lambda_{jk}'$, we now turn our attention to calculating the rate of baryon loss in terms of $\varepsilon_{B\psi}$. In the nuclear matter (n.m.) frame within pulsars, the proper rate of baryon loss via the decay process ${\cal B} \to \psi + \gamma$ is given by~\cite{Berryman:2023rmh}:
\begin{align}
    \dfrac{dn_{\cal B}}{d\tau} = - \frac{\varepsilon_{B\psi}^2 g_{\cal B}^2 e^2}{128\pi^3} (m_{\cal B}^*)^2 \int_{1}^{x_F} & dx \, \sqrt{x^2-1} \times \frac{1+2x\sigma+\sigma^2-\mu^2}{(1+2x\sigma+\sigma^2)^2} \nonumber \\
    & \times \left[ (1+2x\sigma+\sigma^2)(1+x\sigma+2\mu) + \mu^2(1+x\sigma) \right]\,, \label{eq:decay:tot_rate}
\end{align}
where we have defined the dimensionless parameters as follows:
\begin{align}
    x\equiv \frac{E_{\cal B}^{*, \rm{(n.m.)}}}{m_{\cal B}^*}, \quad \quad
    x_F\equiv  \frac{E_{F,{\cal B}}^{*, \rm{(n.m.)}}}{m_{\cal B}^*}, \quad\quad 
    \sigma \equiv \frac{\Sigma_{\cal B}^{\rm{(n.m.)}, 0}}{m_{\cal B}^*}, \quad\quad
    \mu \equiv \frac{m_{\psi}}{m_{\cal B}^*}
   \label{eq:medium:dim_less_def} 
    \,,
\end{align}
in which $E_{\cal B}^{*, \rm{(n.m.)}} = \sqrt{\left(m_{\cal B}^*\right)^2 + |\vec{p}_{\cal B}|^2}$ is related to the total baryon energy \(E_{\cal B}\) by $E_{\cal B} = E_{\cal B}^* + \Sigma_{\cal B}^0$. The quantity $E_{F,{\cal B}}^{*, \rm{(n.m.)}}$ is evaluated on the Fermi surface.

%

\begin{figure}[h]
 \centering
    \includegraphics[width=0.7\linewidth]{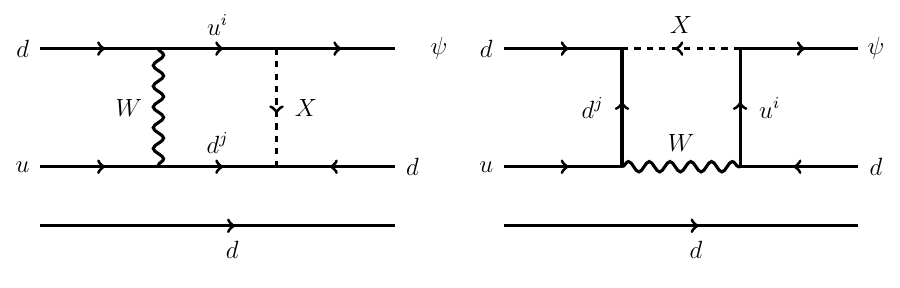}
    \caption{The diagram for $n\to \pi^0 \psi$ decay.}
    \label{fig:ndecay:pion}
\end{figure}

In addition to magnetically induced decays, neutrons and \(\Lambda\) baryons may also decay into mesons, as illustrated in Fig.~\ref{fig:ndecay:pion}. The description of pions, particularly their energy spectra in dense neutron-rich matter, is an active area of research, with recent studies exploring the charged pion mass in such environments~\cite{Fore:2023gwv}. However, the pion energy spectrum at the extreme densities found in neutron star cores remains uncertain. For our analysis, we simplify the evaluation by assuming a constant pion mass, \(m_{\pi^0} = 140\, {\rm MeV}\), to calculate the in-medium decay rate for \(n \to \psi + \pi^0\). Further details are provided in Appendix~\ref{app:dec:pion}, with the resulting limits plotted in Fig.~\ref{fig:pion_lim}. The constraints derived from this pion channel are comparable to those obtained from magnetically induced decays to photons. However, we refrain from combining these results due to the inherent uncertainties in the pion description within dense matter.

In addition to the decay channels previously discussed, baryons can also undergo two-body scatterings, such as $n n \to \psi n$. However, given that we are considering cases where $m_{\psi} \approx m_n$, the final-state neutron is likely to be Pauli blocked, making this scattering process insignificant.  Alternatively, the final state could involve mesons, which are not subject to Pauli blocking, as illustrated in Fig.~\ref{fig:charged_meson_scattering_feynman}. We defer the evaluation of these scattering rates to future work, pending a more reliable understanding of the pion and kaon energy spectra in dense matter.

\begin{table}[hb!]
    \centering
    \begin{tabular}{|c|c|c|c|}
    \hline 
    Name & J0348$+$0432  & J1614$-$2230 & J0737$-$3039A/B\\
        \hline\hline                                          
        $M_p \, (M_{\odot})$ & $2.01(4)$ & $1.908(16)$ & $1.338\, 185(+12, -14)$ [A] \\\hline
        $M_c \, (M_{\odot})$ & $0.172(3)$ & $0.493(3)$ & $1.248\, 868(+13, -11)$ [B] \\\hline
        $|\dot{B}/B|_{2\sigma} \, ({\rm yr^{-1}})$ & $1.8\times 10^{-10}$ & 
        $2.0 \times 10^{-11}$ 
        & $4.0\times 10^{-13}$\\\hline
    \end{tabular}
    \caption{The pulsar-binary systems used for our study: J0348$+$0432~\cite{Antoniadis:2013pzd}, J1614$-$2230~\cite{Arzoumanian:2017puf, NANOGrav:2020gpb}, and J0737$-$3039A/B~\cite{PhysRevX.11.041050}. The corresponding $2\sigma$ limits on the BNV rates, $|\dot{B}/B|_{2\sigma}$ are taken from Ref.~\cite{Berryman:2023rmh}.
    }
    \label{tab:psrbinary}
\end{table}

We translate the limits on the mixing parameter, derived from the observed changes in the orbital period of the three binary systems (see Table~\ref{tab:psrbinary}) in Ref.~\cite{Berryman:2023rmh}, into constraints on the couplings $\lambda_{i} \lambda_{jk}'$ by combining the individual limits on $\varepsilon_{n\psi}$ and $\varepsilon_{\Lambda\psi}$. 

\begin{figure}[ht!]
 \centering
    \includegraphics[width=1\linewidth]{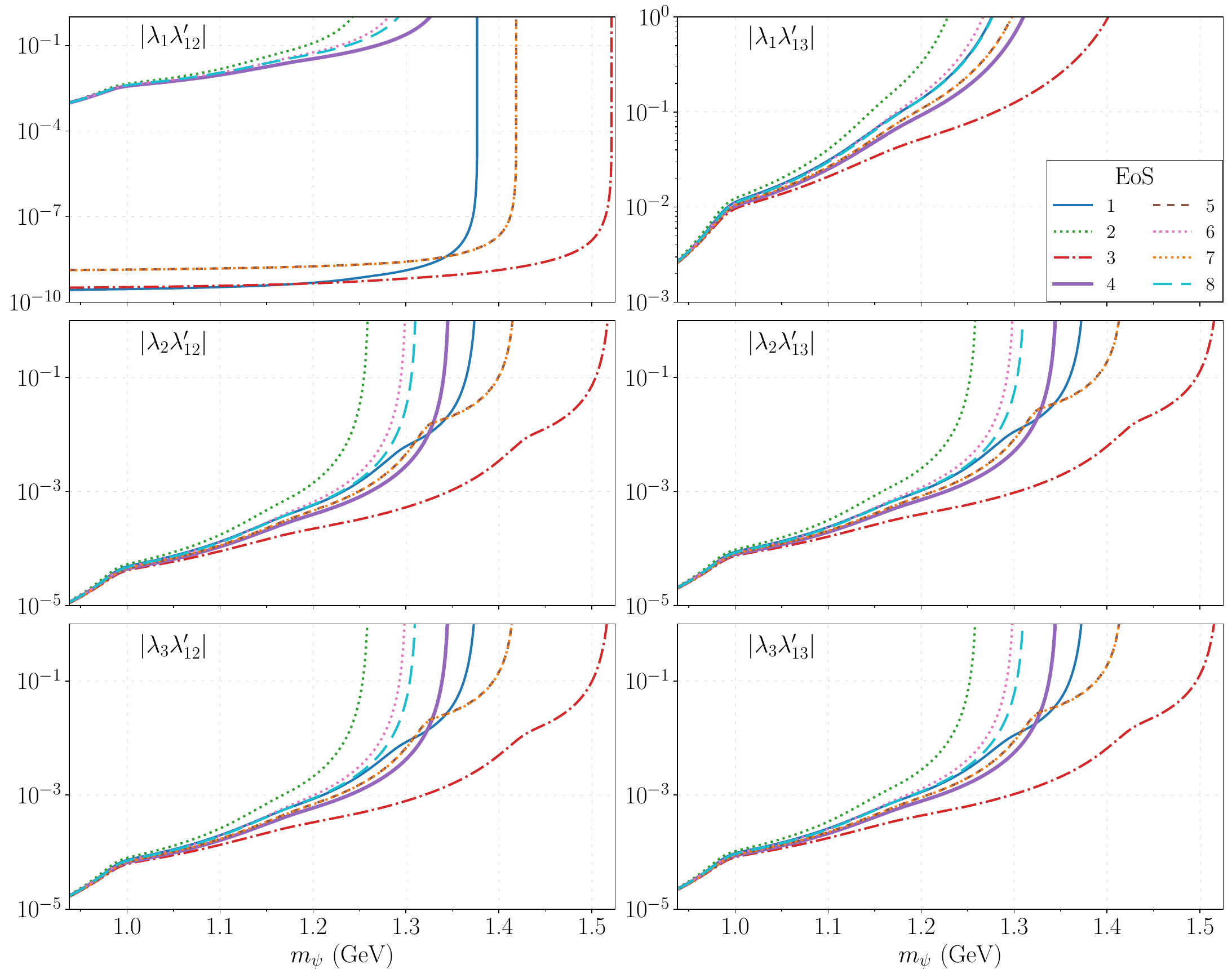}
 \caption{\label{fig:photon_lim}
    Two-$\sigma$ exclusion constraints on the coupling combinations \{$\lambda_{i} \lambda_{12}'$,  $\lambda_{i} \lambda_{13}'$\} for $i \in\{ 1, 2, 3 \}$, derived from pulsar binary limits on baryon decays ${\cal B} \to \psi \gamma$~\cite{Berryman:2023rmh}. The color-charged scalar mass is set at $m_X = 1$ TeV. The figure presents limits across the DS(CMF) EoS family, ranging from DS(CMF)--1 to DS(CMF)--8~\cite{compose_CMF1,compose_CMF8}. The $x-$axis originates at the critical dark baryon mass for nuclear stability, $m_{\psi}^{\rm min} = 937.993$ MeV. The analysis incorporates data from binary pulsar systems PSR J0348+0432, PSR J1614--2230, and PSR J0737--3039A/B.
    }
\end{figure}

To simplify our presentation, the limit on each $\lambda_{i} \lambda_{jk}$ combination is evaluated independently, with the assumption that it alone saturates the total BNV bound, while all other combinations are set to zero. The total BNV rate can be expressed as:
\begin{equation}
\Gamma_{\text{BNV}} = \left(\frac{\varepsilon_{n\psi}}{\bar{\varepsilon}_{n\psi}}\right)^2 \Gamma_n(\bar{\varepsilon}_{n\psi}) + \left(\frac{\varepsilon_{\Lambda\psi}}{\bar{\varepsilon}_{\Lambda\psi}}\right)^2 \Gamma_{\Lambda}(\bar{\varepsilon}_{\Lambda\psi}),
\end{equation}
where $\bar{\varepsilon}_{n\psi}$ and $\bar{\varepsilon}_{\Lambda\psi}$ represent the $2\sigma$ constraints on the mixing parameters when BNV is dominated by neutron and $\Lambda$ decays, respectively~\cite{Berryman:2023rmh}. Thus, the $2\sigma$ limits on $\Gamma_{\text{BNV}}$ is equal to $\Gamma_{\rm BNV}^{2\sigma} = \Gamma_{n}(\bar{\varepsilon}_{n\psi}) = \Gamma_{\Lambda}(\bar{\varepsilon}_{\Lambda\psi})$, and the limits on $\lambda_{i} \lambda_{jk}$ are given by
\begin{equation}
\left| \lambda_{i} \lambda_{jk} \right|^{2\sigma}= \left[ \left(\frac{\tilde{\varepsilon}_{n\psi}^{ijk}}{\bar{\varepsilon}_{n\psi}}\right)^2 + \left(\frac{\tilde{\varepsilon}_{\Lambda\psi}^{ijk}}{\bar{\varepsilon}_{\Lambda\psi}}\right)^2 \right]^{-1/2}, \label{eq:comb:limit}
\end{equation}
where $\tilde{\varepsilon}_{\mathcal{B} \psi}^{ijk} \equiv \varepsilon_{\mathcal{B} \psi} / (\lambda_{i} \lambda_{jk})$. The results are displayed across eight EoS in Fig.~\ref{fig:photon_lim}. The observed kinks in the curves correspond to the threshold energies attainable in each of the three binary systems detailed in Table~\ref{tab:psrbinary}. For lighter $m_{\psi}$ values, particularly at $m_{\psi} \approx 1$ GeV, the various EoS choices yield similar results. However, for heavier $m_{\psi}$, the results vary more significantly. The bounds on combinations of $\lambda_{i} \lambda_{23}'$ are comparatively weaker, attributable to the CKM suppression factor $|V_{ub}|$. The hyperonic EoS selections (\#1, \#7, \#3) facilitate $\Lambda$ decays to $\psi$, which allows us to derive stringent limits on $\lambda_1 \lambda_{12}'$ from tree-level $\Lambda$ decays. It is important to note that even predominantly nucleonic EoS, which initially set hyperon density to zero, may realistically contain some hyperons. This is because unless a complete phase transformation to exotic matter like quarks occurs in the core of neutron stars, a residual population of hyperons, especially $\Lambda$ baryons, is likely. Thus, even a small but non-zero density of hyperons, though lower than in hyperonic EoS, could still lead to significant constraints from tree-level $\Lambda$ decays on $\lambda_1 \lambda_{12}'$.

\begin{figure}[ht!]
 \centering
    \includegraphics[width=\linewidth]{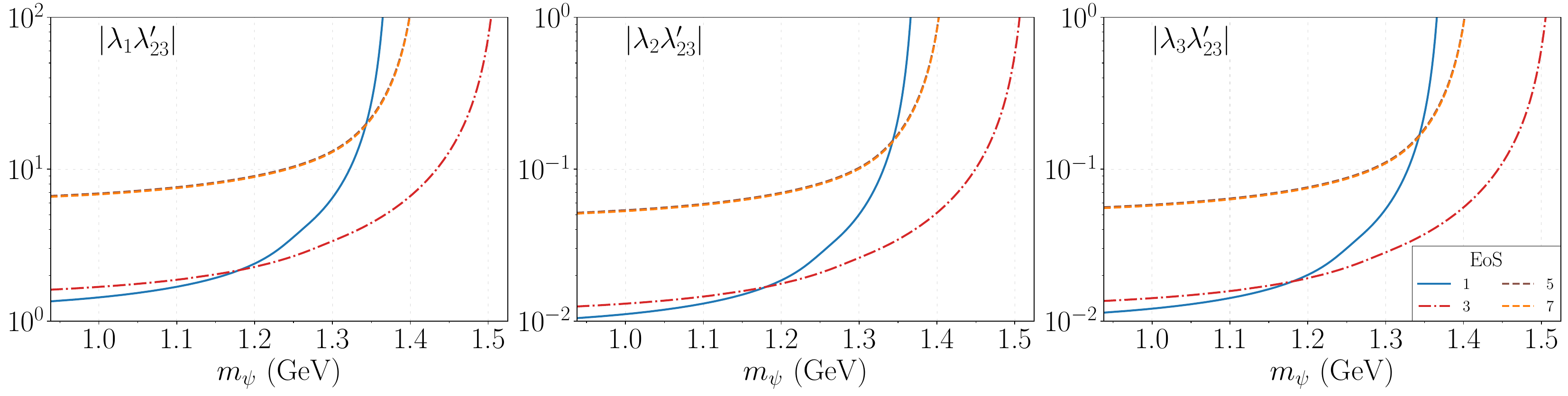}
 \caption{\label{fig:photon_lim:2}
    Two-$\sigma$ exclusion constraints on the coupling combinations \{$\lambda_{i} \lambda_{23}'$\} for $i \in\{ 1, 2, 3 \}$, derived from pulsar binary limits on baryon decays ${\cal B} \to \psi \gamma$~\cite{Berryman:2023rmh}. The color-charged scalar mass is set at $m_X = 1$ TeV. The figure shows constraints derived from the hyperonic variants of the DS(CMF) equation of state family, including DS(CMF)–1, 3, 5, and 7~\cite{compose_CMF1,compose_CMF8}. The $x-$axis originates at the critical dark baryon mass for nuclear stability, $m_{\psi}^{\rm min} = 937.993$ MeV. The analysis incorporates data from binary pulsar systems PSR J0348+0432, PSR J1614--2230, and PSR J0737--3039A/B.
    }
\end{figure}

\section{Discussion and Implications}
\label{sec:discussion}
\begin{figure}[hb!]
\centering\includegraphics[width=0.9\textwidth]{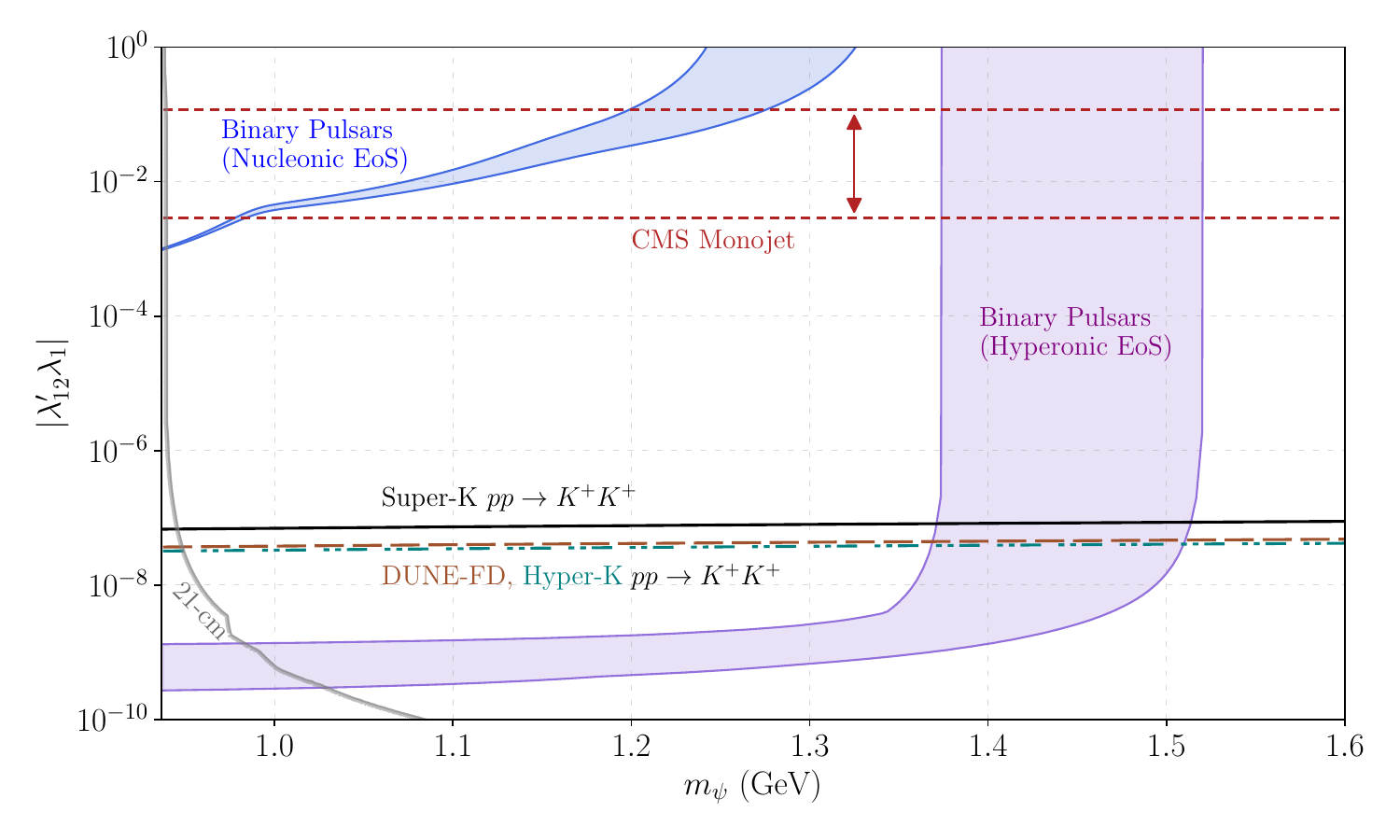}
    \caption{Both laboratory bounds and the bounds derived from binary pulsars in this work are shown together for the $|\lambda_{12}^\prime \lambda_1|$ coupling combination with $m_X = 1$ TeV as a function of the Majorana fermion mass $m_\psi$. The laboratory bounds include the Super-Kamiokande search for $pp \to K^+ K^+$~\cite{sk_kaons} and CMS monojet searches~\cite{Undleeb:2017oor} (95\% C.L.). The CMS monojet constraint places upper bounds on coupling products in the range indicated by the double arrow, depending on the individual choices of $\lambda_{12}^\prime$ and $\lambda_1$ (see text for details). In addition, the projected limits for a $pp \to K^+ K^+$ search at DUNE-FD (dashed) and Hyper-K (dotted) are shown (90\% C.L.). Limits from the pulsar binaries are separated by nucleonic (blue) and hyperonic (purple) EoS, excluding at $2\sigma$ (95.4\% C.L.) all coupling products above their curves. Lastly, we show the 21-cm limits on the lifetime for $\psi \to p \, e^- \, \bar{\nu}$ decays, which could allow for a long-lived DM~\cite{Sun:2023acy} that satisfy the 21-cm measurement below the gray line. The left edge of the plot border corresponds to $m_\psi = m_p - m_e \simeq 937.76$ MeV.}
    \label{fig:lab_and_astro_limits}
\end{figure}

Given the limits on the model parameters from the pulsar binaries we considered in \S~\ref{sec:binary_pulsars} and the laboratory probes in \S~\ref{sec:lab}, a good perspective could be gained by collecting both the terrestrial and astrophysical results together across the couplings we have investigated. The laboratory limits are compared with the exclusions derived in this work from the pulsar binaries in Fig.~\ref{fig:photon_lim} for the $\lambda_{12}^\prime \lambda_1$ coupling combination and $m_X = 1$ TeV, shown in Fig.~\ref{fig:lab_and_astro_limits}. The limits from pulsar binaries are grouped into bands that span the range of either the nucleonic or hyperonic EoS in blue or purple, respectively. Here we see that the searches for rare dinucleon decays ($pp \to K^+ K^+$ in the case of the $|\lambda_{12}^\prime \lambda_1|$ coupling combination) set the most stringent limits above masses $m_\psi > 1.5$ GeV. Utilizing DUNE-FD or Hyper-K as a large volume search for $pp \to K^+ K^+$ could advance limits further still, but within an order of magnitude in the coupling product for the same mediator mass. The bounds from pulsar binaries, on the other hand, that include a modeled EoS with strange quark content can probe even smaller coupling products for the light mass regime, complementing the laboratory bounds. As mentioned previously, uncertainty in the hyperon occupation in the EoS could still yield very strong limits similar to the ones seen here due to the decays of $\Lambda \to \psi \gamma$ at tree level.

One could also relax the bound for $\psi$ to act as a stable DM candidate, $m_\psi \leq m_p + m_e$, by scrutinizing the lifetime estimate of $\psi$ and how it contends with the measurements from 21-cm observations at high redshift. We show this region in Fig.~\ref{fig:lab_and_astro_limits}. This region is determined by ensuring the rate and energy injection of $\psi$ decays satisfy the 21-cm limits on decaying DM~\cite{Sun:2023acy}. For the $\lambda_1 \lambda_{12}^\prime$ coupling combination, the dominant decay mode below $m_\psi < m_\Lambda$ occurs via the three-body decay $\psi \to p \, e^- \, \bar{\nu}$ (mediated by a charged Kaon at first order in the ChPT expansion, see Eq.~\ref{eq:first_order_chpt}), while $\psi \to n \gamma$ is loop-suppressed. The decay rate is given by integrating
\begin{equation}
    d\Gamma =  \frac{1}{32 m_\psi^2 (2\pi)^3} \left(\frac{1}{m_{e\nu}^2-m_K^2}\right)^2 \bigg(\frac{\beta  f_K G_F \lambda_1 \lambda_{12}^\prime}{2 \sqrt{2} f_\pi m_X^2} \bigg)^2 f(m_{e\nu}^2, m_\psi) d m_{e\nu}^2  dE_e
\end{equation}
over the Dalitz variable $m_{e\nu}^2 = (p_e^\mu + p_\nu^\mu)^2$ and the electron energy $E_e$, where $f(m_{e\nu}^2, m_\psi)$ is a phase space factor, $G_F$ is the Fermi constant and $f_K = 156.1$ MeV~\cite{PhysRevD.98.030001} is the kaon decay constant. Since the limits in Ref.~\cite{Sun:2023acy} considered decays of the form $\text{DM} \to e^+ e^-$, the visible energy in the 2-body final state is assumed as the dark matter mass $m_\text{DM}$. Since the 3-body decay yields an average final state lepton energy of $(m_\psi - m_p)/2$, we can determine the 21-cm limits approximately by matching to the effective mass $m_\text{DM}^\text{eff} \sim (m_\psi - m_p)/2$ and dividing the 95~\% C.L. lifetime constraint by $m_\psi / m_\text{DM}^\text{eff}$ to compensate for the larger DM number density that would result from the smaller mass $m_\text{DM}^\text{eff}$. The dashed gray line in the bottom corner of Fig.~\ref{fig:lab_and_astro_limits} shows that to satisfy the 21-cm limits on the lifetime of $\tau \gtrsim 10^{28}$ s permits $m_\psi > m_p$ as long as the coupling product is well below $|\lambda_1 \lambda_{12}^\prime| < 10^{-7}$. As more phase space for the decay opens up at heavier masses, this limit is pushed to the $10^{-10}$ level, but within the mass range $m_p + m_e < m_\psi < m_\Lambda$ the hyperonic EoS binary pulsar limits are able to probe the stable DM solution.

\begin{figure}[h!]
    \centering
    \includegraphics[width=0.49\textwidth]{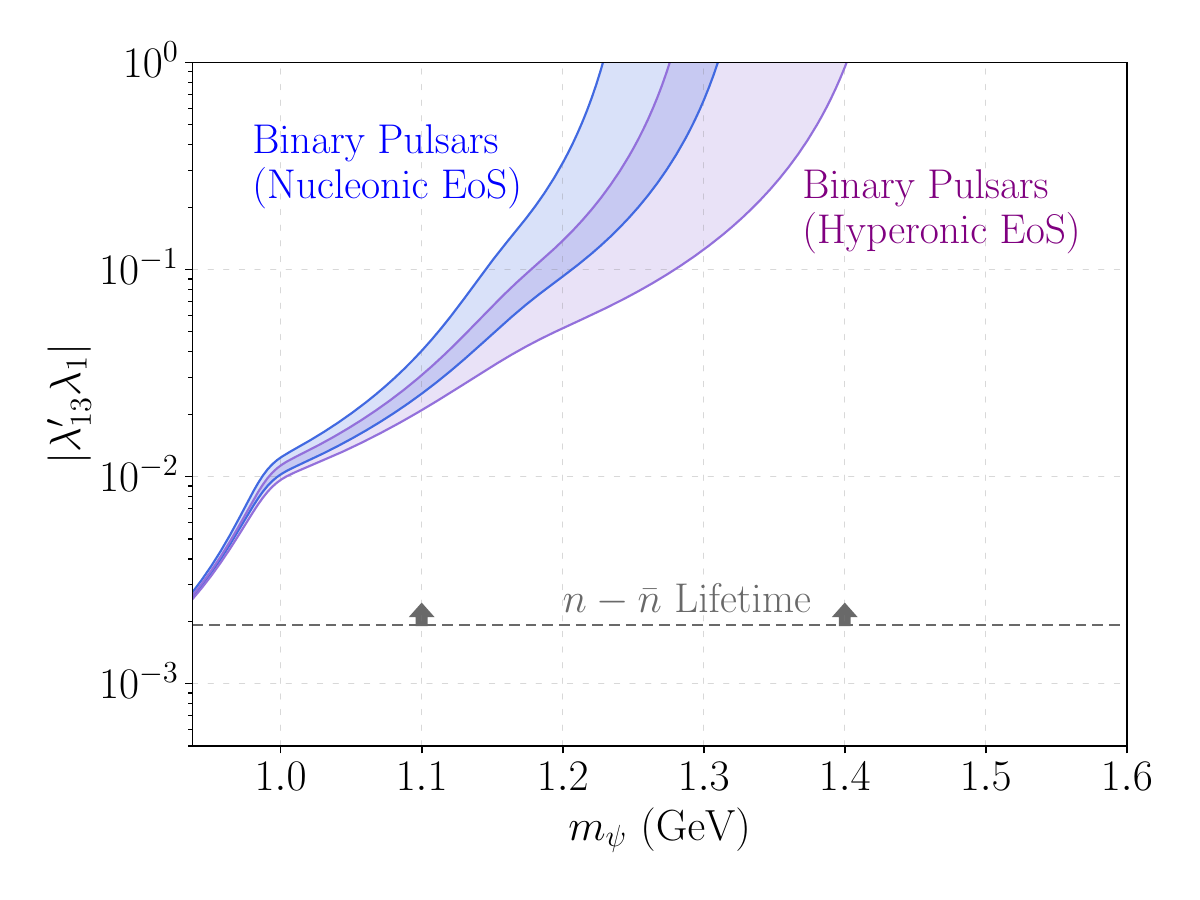}
    \includegraphics[width=0.49\textwidth]{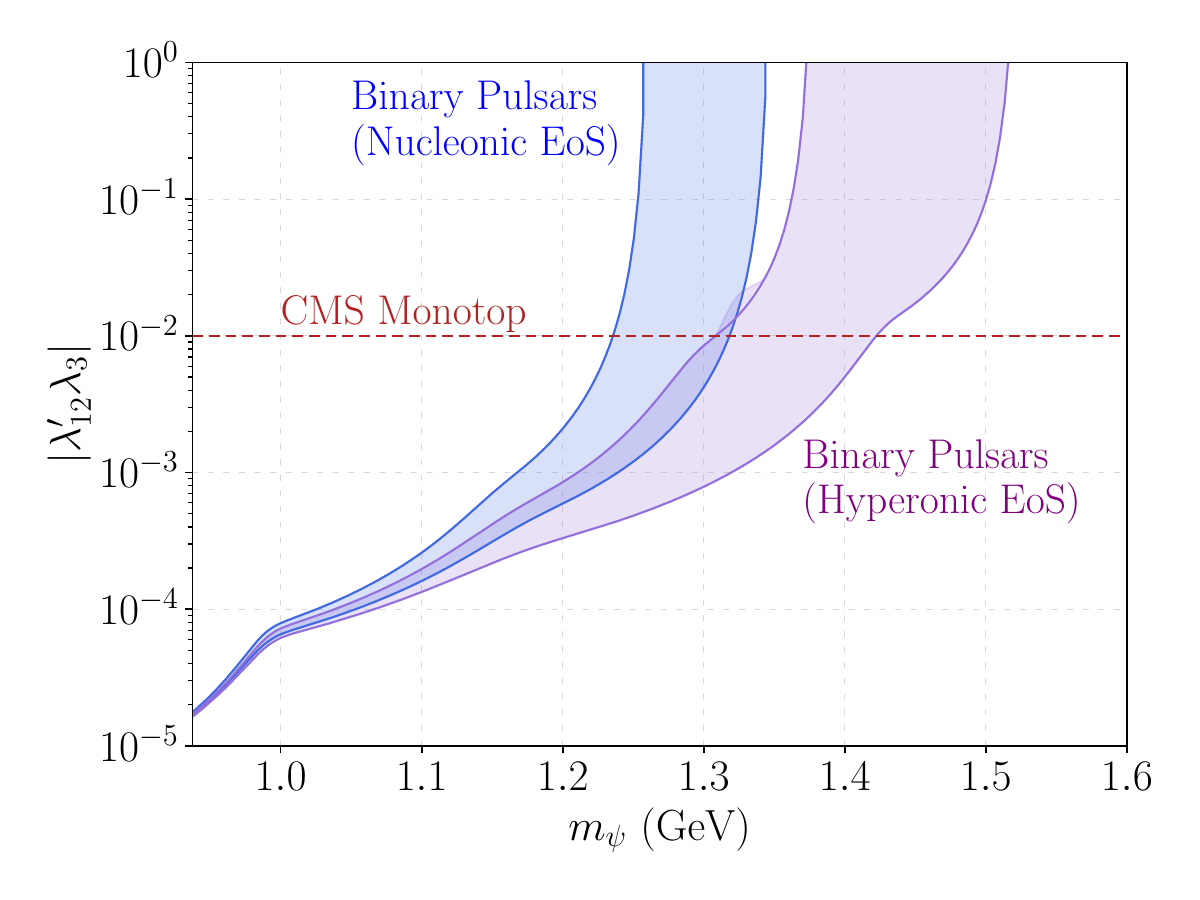}
    \caption{We show laboratory bounds and the bounds derived from binary pulsars (this work) together for the $|\lambda_{13}^\prime \lambda_1|$ coupling combination (left) and $|\lambda_{12}^\prime \lambda_3|$ (right) with $m_X = 1$ TeV as a function of the Majorana fermion mass $m_\psi$. As in Fig.~\ref{fig:lab_and_astro_limits}, limits from the pulsar binaries are separated by nucleonic (blue) and hyperonic (purple) EoS, excluding at 
    $2-\sigma$ (95.4\% C.L.) all coupling products above their curves. Here the limits from dinucleon decay $pp \to K^+ K^+$ do not appear due to large CKM and loop factor suppression. In the case of $|\lambda_{13}^\prime \lambda_1|$, the $n-\bar{n}$ lifetime constraint (90~\% C.L.) already rules out coupling combinations greater than $\sim 2 \times 10^{-3}$~\cite{PhysRevD.103.012008}. For $|\lambda_{12}^\prime \lambda_3|$, bounds from the monotop search at CMS (95~\% C.L.) are shown~\cite{Sirunyan:2018gka}.}
    \label{fig:lab_and_astro_limits_lam12_lam3}
\end{figure}

Let us examine the model parameter space for higher-generational couplings. For example, in Fig.~\ref{fig:lab_and_astro_limits_lam12_lam3}, left, we show the binary pulsar limits on the $|\lambda_{13}^\prime \lambda_1|$ coupling combination which necessitates a $d-b$ CKM element at 1-loop to facilitate $n \to \psi \gamma$ decay. With this same coupling combination we also generate the dim-9 operator responsible for $n-\bar{n}$ oscillations discussed in \S~\ref{sec:nnbar}; the limits on the neutron oscillation lifetime are shown in green. Since the $n-\bar{n}$ lifetime measurements already exclude parameter space beyond the sensitivity of binary pulsars, it is likely that future searches for neutron oscillations will be the best channel to probe deeper in the coupling. For example, a goal of three orders of magnitude improvement in the sensitivity to the oscillation lifetime -- nearing $10^{11}$ seconds -- has been set by the NNBAR collaboration~\cite{Backman:2022szk}. This translates to an improvement in sensitivity to 10 times smaller couplings. As for limits from di-proton decay, there are no competitive limits in this coupling combination, since it would suffer from a 2-loop suppression and suppression from CKM matrix elements.

In Fig.~\ref{fig:lab_and_astro_limits_lam12_lam3}, right, the parameter space for top quark and down/strange quark couplings via $|\lambda_{12}^\prime \lambda_3|$ are shown. These contrast the laboratory searches for monotop final states that arise from down-strange fusion at the LHC~\cite{Sirunyan:2018gka} with the binary pulsar limits. As was the case in Fig.~\ref{fig:lab_and_astro_limits}, the monotop searches test a region of parameter space for larger fermion masses where $\psi$ is not a stable DM candidate, but where the $X$ decays (and potentially $\psi$ decays, e.g.~\cite{Davoudiasl:2015jja}) could still seed the baryon asymmetry of the universe. There are no other laboratory-based probes of this coupling combination to-date (as with Fig.~\ref{fig:lab_and_astro_limits_lam12_lam3}, left, the di-proton decay limits are heavily CKM- and loop-suppressed), leaving the limits from binary pulsar orbital period decay as the most stringent down to the level of $\sim10^{-5}$ in the coupling product at $m_X = 1$ TeV.

One could also ask what region of parameter space could allow $\psi$ to be a stable DM candidate (at least satisfying the 21-cm limits) for the coupling combinations in Fig.~\ref{fig:lab_and_astro_limits_lam12_lam3}. The decays for $\lambda_{13}^\prime \lambda_1$ could proceed through $\psi \to p e^- \nu$ via a virtual $B$ meson, for instance, which may be somewhat suppressed in comparison to the kaon-mediated decay for $\lambda_{12}^\prime \lambda_1$, but still quite strong. No simple tree-level decay exists for $\lambda_{12}^\prime \lambda_3$ which requires a top quark, but both coupling combinations permit $\psi \to n \gamma$ at 1-loop. We find that the $\psi \to n \gamma$ vacuum lifetime satisfies the 21-cm limits for coupling products $|\lambda_{13}^\prime \lambda_1| \lesssim 10^{-5}$ and $|\lambda_{12}^\prime \lambda_3| \lesssim 10^{-8}$. Each of these limits are well out of reach of the existing binary pulsar and terrestrial sensitivities, except of course within kinematically forbidden range $m_\psi \leq m_p + m_e$. 

\section{Conclusion}
\label{sec:conclusion}

In this paper, we have explored the effect of BNV interactions in neutron stars by studying a minimal extension of the SM 
This model involves a TeV-scale colored scalar mediator and a GeV-scale Majorana fermion $\psi$ (which can also act as the dark DM candidate). The $\Delta B = 1$ effective interactions between the SM quarks and the Majorana fermion in the model translate into a mass loss inside neutron stars, which is detectable in binary pulsar systems through precise measurements of their orbital period. 

The decay of baryons $\mathcal{B} \to \gamma \psi$ (for $\mathcal{B} = n, \Lambda$) is followed by the scattering into meson final states such as $\psi n \to \pi^- K^+$. This amounts to net $\Delta B = 2$ processes in the star that precipitates a mass loss that can be observable in the orbital period of the star in binary systems. Since the scattering rate happens at an equal or faster rate than the decay channels, no significant build-up of a $\psi$ population is formed. By comparing the predicted baryon number rate of change with measurements on the orbital period decay of the binary pulsars, we find strong constraints on the model parameter space. They are particularly stringent when considering hyperonic equations of state that allow for enhanced decay rates at tree-level from $\Lambda \to \psi \gamma$ decays. 

The binary pulsar systems we consider exclude coupling products of the lowest generational combination ($|\lambda_1 \lambda_{12}^\prime|$) as low as $\sim 10^{-9}$ for a 1 TeV mediator mass; instead 
$|\lambda_1 \lambda_{12}^\prime| \sim \mathcal{O}(1)$ translates to an exclusion reach up to $\mathcal{O}(10^4)$ TeV in the mediator mass. For higher-generational couplings the CKM suppression becomes large and the sensitivity is much weaker resulting in exclusions that range from $\sim 10^{-5}$ to $10^{-3}$ (except for those involving $\lambda_{23}^\prime$).
LHC searches via monotop, monojet, and dijet final states are well-motivated and may have better sensitivity over weakly constrained parameter space in these cases. More intense searches for $n-\bar{n}$ oscillation are also motivated here, which we find to already be more sensitive than limits from binary pulsars on the relevant higher-dimensional operators. 

We also investigated the possibility of direct detection, when $\psi$ is the DM candidate, through the scattering channel $\psi n \to \pi^- K^+$ via the $\lambda_1 \lambda_{12}^\prime$ combination in a large detector volume like those at DUNE-FD or Hyper-K. This leaves a very unique final state signature for a direct detection search, although resulting limits are not as strong as those from the binary pulsars for hyperonic equation of state. 
Still, one could imagine similar scattering processes for other flavor combinations beyond $uds$; for example, $\psi p \rightarrow \ell^+ \nu_\ell$ or $\psi p \rightarrow \pi^0 \ell^+ \nu_\ell$ via a virtual $B$-meson. In such a situation, the language of the 3-flavor chiral perturbation expansion is no longer sufficient, and one must use other methods. We leave this possibility to a future work.

Another interesting direction for future investigation is to consider a model where $\psi$ has a Dirac nature.
In this case the constraints from double proton decay disappear as the $p p \to K^+ K^+$ process relies on the Majorana mass of $\psi$. 
The most stringent bound on $| \lambda_1 \lambda_{12}^\prime |$ would then be set by $\Lambda \to \bar{\psi} \gamma$ decay in neutron stars 
where $\bar{\psi}$ remains inside the star without an $\bar{n}$ presence. In this case, the binary pulsar orbital period would still set strong constraints, but they would originate instead from a combination of effects, notably changes to the equation of state due to a population of $\bar{\psi}$ gravitationally trapped within the star.  
The production and subsequent trapping of feebly interacting particles in the neutron star interior is also an interesting subject for a future study.

While we studied a specific model of BNV that permits these decay and scattering channels, other similar models should in principle face similar constraints to the ones we derive here from binary pulsar systems. These constraints may limit the parameter space of viable baryogenesis depending on the coupling structure of the specific model, and indeed, among the families of baryogenesis-inspired EFTs, many of them predict mixings between the SM baryons and particles in the dark sector. The presence of these mixings to facilitate baryon decays and mass loss in neutron stars should then be considered in the corresponding model parameter space 
in a similar spirit to the constraints we have laid out here.

\acknowledgments
The work of R.A. is supported in part by NSF Grant No. PHY-2210367. M.Z. acknowledges partial support from the U.S. Department of Energy (DOE) under contract DE-FG02-96ER40989. A.T. acknowledges support in part by the DOE grant DE-SC0010143. We are grateful to the Center for Theoretical Underground Physics and Related Areas (CETUP*), the Institute for Underground Science at Sanford Underground Research Facility (SURF), and the South Dakota Science and Technology Authority for their hospitality and financial support. Their stimulating environment was invaluable during the period in which this work started. R.A. wishes to thank Center for 
Cosmology and AstroParticle Physics (CCAPP) at The Ohio State University for their kind hospitality, and John Beacom for useful conversations, while part of this work was completed.

\bibliography{main.bib}

\appendix

\section{Expansion of the new physics Lagrangian:}
\label{app:chiral:expansion}

We are interested in expanding the new physics term given in
Eq.~\eqref{eq:chiPT_lagr},
\begin{equation}
\label{eq:chiPT_lagr_appendix}
    \mathcal{L}_\text{eff,ChPT}^{(0)} = \beta \text{Tr}\bigg[ \hat{C}^R u^\dagger B_R u\bigg]\psi_R + \text{h.c.}
\end{equation}
which was obtained by matching the quark-level operator $(d s) u \psi$ to the chiral perturbation theory using the spurion
\begin{align}
    \hat{C} [(d s)u] = \dfrac{\lambda_{1}\lambda^\prime_{12}}{m_X^2}\begin{bmatrix}
        1 & 0 & 0 \\
        0 & 0 & 0 \\
        0 & 0 & 0
    \end{bmatrix}.
\end{align}
Expanding 
Eq.~\eqref{eq:chiPT_lagr_appendix} in $1/f_\pi$ to zeroth order, we have
\begin{equation}
\label{eq:zeroth_order_chpt}
    \mathcal{L}_\text{eff,ChPT}^{(0)} =  \beta\dfrac{ \lambda_{1}\lambda^\prime_{12}}{m_X^2} \bigg( \frac{1}{\sqrt{6}}\bar{\psi}^c P_R \Lambda+\frac{1}{\sqrt{2}}\bar{\psi}^c P_R \Sigma^0 + \text{h.c.} \bigg) + \mathcal{O}(1/f_\pi)
\end{equation}
These two terms give rise to a mass mixing between the hyperons and $\psi$ which is suppressed by $1/m_X^2$ which is order TeV$^{-2}$. It also allows for the ``pole'' term in Fig.~\ref{fig:charged_meson_scattering_feynman}, middle.

To first order in $1/f_\pi$, we take $u^\dagger \simeq 1 - i \frac{\phi}{2f_\pi}$ and keep terms linear in $1/f_\pi$;
\begin{align}
\label{eq:first_order_chpt}
    \mathcal{L}_\text{eff,ChPT}^{(0)} \supset  \frac{\beta}{f_\pi}\dfrac{\lambda_{1}\lambda^\prime_{12}}{m_X^2} \bigg(& \frac{i K^- \bar{\psi}^c P_R p }{\sqrt{2} } -\frac{i K^+ \bar{\psi}^c P_R\Xi^- }{\sqrt{2} }+\frac{i \pi^- \bar{\psi}^c P_R\Sigma^+ }{\sqrt{2} }-\frac{i \pi^+ \bar{\psi}^c P_R \Sigma^- }{\sqrt{2} } + \text{h.c.} \bigg) + \mathcal{O}(1/f_\pi^2)
\end{align}
To order $1/f_\pi^2$, we take $u^\dagger \simeq 1 - i \frac{\phi}{2f_\pi} - \frac{\phi^\dagger \phi}{8 f_\pi^2} $ and keep terms $\mathcal{O}(1/f_\pi^2)$;
\begin{align}
    \mathcal{L}_\text{eff,ChPT}^{(0)} \supset \frac{\beta}{f_\pi^2}\dfrac{\lambda_{1}\lambda^\prime_{12}}{m_X^2} \bigg(
    &
    -\sqrt{\frac{3}{8}} K^- K^+ \bar{\psi}^c P_R \Lambda -\frac{K^- K^+ \bar{\psi}^c P_R \Sigma^0 }{2 \sqrt{2}} + \frac{{\pi^+} K^- \bar{\psi}^c P_R n }{2}+\sqrt{\frac{3}{2}}\frac{{\eta^8} K^- \bar{\psi}^c P_R  p}{4  } 
    +\frac{\pi^0 K^- \bar{\psi}^c P_R  p }{4 \sqrt{2} } \nonumber \\
    &+\sqrt{\frac{3}{2}}\frac{{\eta^8} K^+ \bar{\psi}^c P_R \Xi^-}{4} + \frac{{\pi^-} K^+ \bar{\psi}^c P_R\Xi^0 }{2 }
    +\frac{\pi^0 K^+ \bar{\psi}^c P_R\Xi^- }{4 \sqrt{2} } -\frac{K^0 \pi^+  \bar{\psi}^c P_R \Xi^- }{4 F^2} \nonumber \\
    & +\frac{{\pi^0} {\pi^-} \bar{\psi}^c P_R \Sigma^+}{2 \sqrt{2} } 
    +\frac{{\pi^0} {\pi^+} \bar{\psi}^c P_R \Sigma^-}{2 \sqrt{2} } -\frac{K^+ \overline{K^0} \bar{\psi}^c P_R\Sigma^- }{4 } -\frac{\pi^- \overline{K^0} \bar{\psi}^c P_R p}{4}  \nonumber \\
    &-\frac{K^0 K^-  \bar{\psi}^c P_R \Sigma^+}{4} -\frac{1}{\sqrt{2}} \pi^- \pi^+ \bar{\psi}^c P_R \Sigma^0 + \text{h.c.} \bigg) + \mathcal{O}(1/f_\pi^3)
\end{align}
For loop-level mixings generated via Fig.~\ref{fig:n:mix} in Eq.~\ref{eq:n_mixing} and Eq.~\ref{eq:lambda_mixing:loop}, the spurions include the loop factors as
\begin{align}
\label{eq:lam_mix_spurion}
    \hat{C}^R[(ds)u] &= 
    \frac{G_F \sqrt{3}}{8\pi^2 m_W^2}
    \sum_{i=1}^3 \lambda_i  m_{u_i} \left[ \sum_{j= 2}^{3}  \lambda_{1j}' V_{i 2} + \sum_{j= 1, 3} \lambda_{2j}' V_{i 1} \right] V^{*}_{1 j} m_{d_j}\,  F(x_{d_j}, x_{u_i}, x_{X}) \times
    \begin{pmatrix}
1 & 0 & 0\\
0 & 0 & 0\\
0 & 0 & 0
\end{pmatrix}
\end{align}
\begin{align}
\label{eq:n_mix_spurion}
    \hat{C}^R[(dd)u] &= 
    \frac{G_F}{\sqrt{2} \, 8\pi^2 m_W^2}
    \sum_{i} \sum_{j\neq 1} \lambda_{i} \lambda_{1j}' V_{i1} V^{*}_{1j} m_{d_j} m_{u_i}\,  F(x_{d_j}, x_{u_i}, x_{X}) \times
    \begin{pmatrix}
0 & 0 & 0\\
0 & 0 & 0\\
0 & 1 & 0
\end{pmatrix}
\end{align}
In Eq.~\ref{eq:lam_mix_spurion}, the spurion texture is exactly the same as the tree-level mixing to hyperons, but the prefactor is changed to include the loop factors. In Eq.~\ref{eq:n_mix_spurion}, the different texture gives rise to a different ChPT expansion.

\section{Scattering Amplitudes}
\label{app:scattering}
The scattering amplitude for the process $\psi n \to K^+ \pi^-$ is dominated by the center diagram as shown in  Fig.~\ref{fig:charged_meson_scattering_feynman}. This occurs via the zeroth-order mass mixing terms in $\mathcal{L}_\text{eff,ChPT}^{(0)}$ in 
Eq.~\eqref{eq:zeroth_order_chpt}. Then, together with the operator at order $1/f_\pi^2$,
\begin{align}
    \mathcal{L}_{\phi B}^{(1)} &= \frac{D}{2} {\rm Tr} \left( \overline{B} \gamma^{\mu} \gamma_{5} \{u_{\mu}, B\}\right) +  \frac{F}{2} {\rm Tr} \left( \overline{B} \gamma^{\mu} \gamma_{5} [u_{\mu}, B]\right) \nonumber \\
    &\supset \dfrac{(D+3F)}{f_\pi^2} \bigg[ \frac{i}{4\sqrt{3}} \bigg( (\overline{K^0} \partial_\mu \pi^0) \overline{\Lambda^0}\gamma^\mu \gamma^5 n  + (\pi^0 \partial_\mu K^0) \overline{n}\gamma^\mu \gamma^5 \Lambda^0 \bigg) \nonumber \\
    & - \frac{i}{2\sqrt{6}} \bigg( (\pi^- \partial_\mu K^+) \overline{n}\gamma^\mu \gamma^5 \Lambda^0 + (K^-\partial_\mu \pi^+) \overline{\Lambda^0}\gamma^\mu \gamma^5 n \bigg)\nonumber\bigg]
\end{align}
the 4-point contact interaction between the neutral hyperons and the pions and kaons lends to the middle diagram in Fig.~\ref{fig:charged_meson_scattering_feynman}. 
The amplitude for this process is
\begin{align}
  i \mathcal{M} &=  \dfrac{(D+3F)}{f_\pi^2} \beta  \dfrac{\lambda_{1}\lambda^\prime_{12}}{m_X^2} \frac{i}{4\sqrt{3}}  \braket{K^0 \pi^0 \mid \big( (\overline{K^0} \partial_\mu \pi^0) \big[ \frac{1}{\sqrt{6}}\bar{\psi}^c P_R \Lambda^0 \big]\overline{\Lambda^0}\gamma^\mu \gamma^5 n \big) \mid n \psi } \nonumber \\
  &= A \braket{K^0 \pi^0 \mid  (\overline{K^0} \partial_\mu \pi^0)  \bar{\psi}^c P_R \mid 0} \braket{0 \mid \Lambda^0  \overline{\Lambda^0} \mid 0} \braket{0 \mid \gamma^\mu \gamma^5 n  \mid n \psi } \nonumber \\
  & \begin{aligned}=A \bra{0} a_k a_{p_\pi} \sqrt{4 E_\pi E_K} &\int d^4 x \int d^4 y (\overline{K^0} \partial_\mu \pi^0)_x \bar{\psi}^c (y) P_R \int \frac{d^4 q}{(2\pi)^4} \frac{-i(\slashed{q} + m_\Lambda) e^{-iq(x-y)}}{q^2 - m_\Lambda^2} \nonumber \\
  & \times \gamma^\mu \gamma^5 n(x) \sqrt{4 E_n E_\psi} a^\dagger_{p_\psi} a^\dagger_{p_n} \ket{0}
    \end{aligned}\nonumber \\
  &= A \int d^4 x d^4 y \frac{d^4 q}{(2\pi)^4} p_{\pi, \mu} [\bar{u}_{p_\psi} \frac{1 + \gamma^5}{2} \gamma^\mu \gamma^5 u_{p_n}] \frac{-i(\slashed{q} + m_\Lambda)}{q^2 - m_\Lambda^2} e^{-i p_\psi y} e^{i k x} e^{i p_\pi x} e^{-i p_n x} e^{-iq(x-y)} \nonumber \\
  &= A \int d^4 x d^4 q \, \, p_{\pi, \mu} [\bar{u}_{p_\psi} \frac{1 + \gamma^5}{2} \gamma^\mu \gamma^5 u_{p_n}] \frac{-i(\slashed{q} + m_\Lambda)}{q^2 - m_\Lambda^2} \delta^4 (p_\psi - q)  e^{i (p_\pi + k - p_n - q)x} \nonumber \\
  &= (2\pi)^4 A p_{\pi, \mu} [\bar{u}_{p_\psi} \frac{1 + \gamma^5}{2} \gamma^\mu \gamma^5 u_{p_n}] \frac{-i(\slashed{p}_\psi + m_\Lambda)}{p_\psi^2 - m_\Lambda^2}  \delta^4 (p_\pi + k - p_n - q)
\end{align}
where $A \equiv \dfrac{(D+3F)}{f_\pi^2} \beta  \dfrac{\lambda_{1}\lambda^\prime_{12}}{m_X^2} \frac{i}{4\sqrt{18}}$ and we suppress the spin indices. Squaring, summing over spins, and averaging over initial spin states gives
\begin{equation}
    \braket{\mathcal{M}}^2 = \frac{1}{4} A^2 \frac{(m_\Lambda+m_\psi)^2\big[m_n^2 m_\pi^2 - 2 m_n m_\pi^2 m_\psi + m_\pi^2 t - (m_\psi^2-t) (m_\psi^2-s-t)) - m_K^2 (m_\pi^2+m_\psi^2-t))\big]}{12 (m_\Lambda^2-m_\psi^2)^2}
\end{equation}
A similar matrix element follows for the $\Sigma^0 -\psi$ mixing. 

\section{Decay to Pions in Neutron Stars}
\label{app:dec:pion}
In this section, we extend the Chiral Perturbation Theory (ChPT) formalism from Ref.~\cite{Alonso-Alvarez:2021oaj} to incorporate medium effects in the evaluation of \(n \to \psi \pi^0\) decays within the cores of neutron stars, using Relativistic Mean Field (RMF) theory. The in-medium amplitude for the \(n \to \psi \pi^0\) decay, as illustrated in Fig.~\ref{fig:ndecay:pion}, is given by
\begin{align}
    \mathcal{M}_m = 
    &\frac{i G_F}{\sqrt{2}\, 8\pi^2\, m_W^2}
    \sum_{i} \sum_{j\neq 1} \lambda_{i} \lambda_{1j}' V_{i1} V^{*}_{1j} m_{d_j} m_{u_i}\,  F(x_{d_j}, x_{u_i}, x_{X}) \\
    &\qquad \qquad \qquad \qquad \times \bar{u}_{\psi} P_R \left[ W_{n0}^L(q^{*2}) - \slashed{q}^* W_{n1}^L(q^{*2})\right] u_n,
\end{align}
where $q^{*}_{\mu} \equiv q_{\mu} - \Sigma_{\mu}$, and there are two contributions to the form factors $W_{n0}^L$. The first comes from the baryon-pole contributions to $n \to \pi n$ as discussed in Ref.~\cite{Alonso-Alvarez:2021oaj}, based on Eq.~\eqref{eq:chpt:Lagr}, where we adopt the values $D = 0.80$ and $F = 0.46$~\cite{Ledwig:2014rfa}. The second contribution arises from the contact term diagram for $n \to \pi$. Altogether, we obtain:
\begin{equation}
\begin{split}
    W_{n 0}^{L, {\rm pole}}(q^{*2}) =& \left(\frac{-\beta b_n}{f} \right) \frac{q^{*2} + m_n^{*2}}{m^{*2}_n - q^{*2} } - \frac{\beta}{f} c_n,\\
    W_{n 1}^{L, {\rm pole}}(q^{*2}) =& \left(\frac{-\beta b_n}{f} \right) \frac{2 m^*_n}{m^{*2}_n - q^{*2} },
\end{split}
\end{equation}
where $b_n = -(D+F)/2$, and $c_n=0.5$.


Assuming one combination of \((i, j)\) couplings at a time, the spin-averaged squared amplitude is given by
\begin{align}
    \overline{\left|\mathcal{M}_m\right|^2} = 
    &\frac{G_F^2}{256\pi^4\, m_W^4}
   |\lambda_{i} \lambda_{1j}' V_{i1} V^{*}_{1j}|^2 m_{d_j}^2 m_{u_i}^2\,  \left[ F(x_{d_j}, x_{u_i}, x_{X}) \right]^2 \times \left\{ A\right\},
\end{align}
where
\begin{equation}
    \begin{split}
    A \equiv& {\rm Tr} \left\{\left[\slashed{q} + m_{\psi}\right] P_R\left[ W_{n0}^L(q^{*2}) - \slashed{q}^* W_{n1}^L(q^{*2})\right] \left[\slashed{p}^* + m_{n}^*\right] \left[ W_{n0}^L(q^{*2}) - \slashed{q}^* W_{n1}^L(q^{*2})\right] P_L\right\}\\
    =& {\rm Tr} \left\{P_L \slashed{q} \left[ W_{n0}^L(q^{*2}) - \slashed{q}^* W_{n1}^L(q^{*2})\right] \left[\slashed{p}^* + m_{n}^*\right] \left[ W_{n0}^L(q^{*2}) - \slashed{q}^* W_{n1}^L(q^{*2})\right]\right\}\\
    =& 2 W_1^2 \left[ 2 (p^* \cdot q^*) (q \cdot q^*) - (q^*)^2 (q\cdot p^*) \right] + 2 W_0^2 (q\cdot p^*) - 4 W_0 W_1 m_n^* (q\cdot q^*),
    \end{split}
\end{equation}
which, in the vacuum limit, simplifies to
\begin{equation}
    \begin{split}
    A_{\rm vac} =& 2 (p \cdot q)\left[ m_{\psi}^2  W_1^2 + W_0^2 \right] - 4 W_0 W_1 m_n m_{\psi}^2,
    \end{split}
\end{equation}
and further reduces to:
\begin{equation}
    \begin{split}
    A_{\rm vac}^{\rm cm} =& 2 m_n \left[ E_{\psi}^{\rm cm}\left(  W_0^2 + m_{\psi}^2  W_1^2 \right) - 2 W_0 W_1 m_{\psi}^2\right],
    \end{split}
\end{equation}
in the neutron's cm frame in vacuum, consistent with the expression in Eq.~(32) of Ref.\cite{Alonso-Alvarez:2021oaj}. 

The decay rate for $n \to \pi^0 \psi$ in the neutron's cm frame is given by:
\begin{align}
    \Gamma_{{\rm cm}}({\vec{p}}_{n}^{\, \rm nm}) =& \frac{1}{32\pi^2} \frac{|\vec{q}^{\, {\rm cm}}| }{E^{*, {\rm cm}}_n\, E_n^{{\rm cm}}} \, \int d\Omega_{q}\, \overline{|\mathcal{M}_m|^2},
\end{align}
where $\vec{p}_{n}^{\, \rm nm}$ is the neutron's momentum in the nm frame, and $\vec{q}^{\, {\rm cm}}$ is the momentum of $\psi$ in the neutron's cm frame. This rate is then boosted to the nm frame using the Lorentz factor $\gamma = \left( 1 - (\vec{p}_n^{\, \rm nm}/E_n^{\rm nm})^2\right)^{-1/2}$. The decay rate in the nm frame is integrated over the neutron star's volume using Eq.~\eqref{eq:B:rate} to compute the baryon-loss rate.

\begin{figure}[!t]
 \centering
    \includegraphics[width=0.9\linewidth]{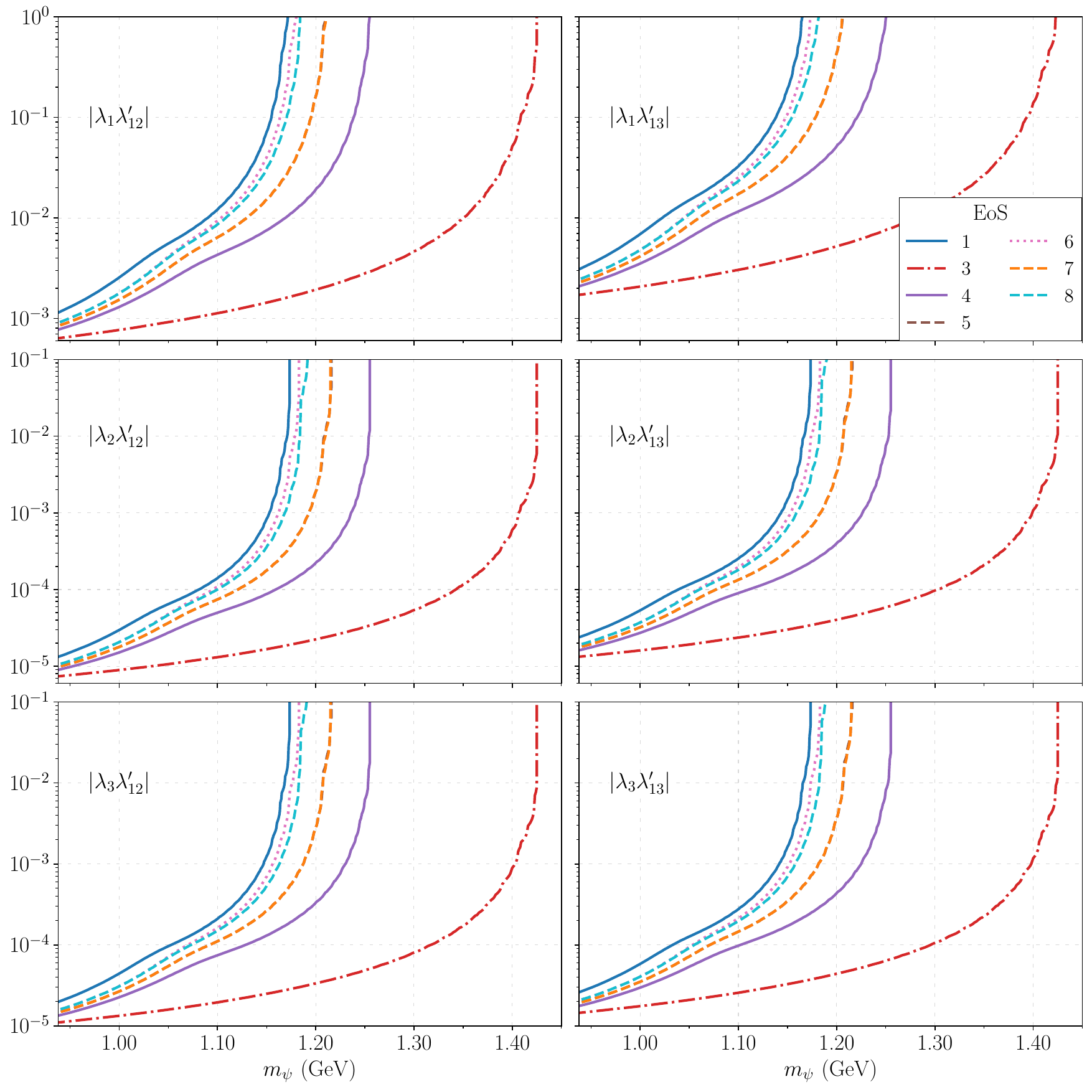}
 \caption{\label{fig:pion_lim}
    Two-$\sigma$ exclusion constraints on the coupling combinations \{$\lambda_{i} \lambda_{12}'$,  $\lambda_{i} \lambda_{13}'$\} for $i \in\{ 1, 2, 3 \}$, derived from pulsar binary limits on baryon decays $n \to \psi \pi^0$. The color-charged scalar mass is set at $m_X = 1$ TeV. The figure presents limits across the DS(CMF) EoS family, ranging from DS(CMF)--1 to DS(CMF)--8~\cite{compose_CMF1,compose_CMF8}. The $x-$axis originates at the critical dark baryon mass for nuclear stability, $m_{\psi}^{\rm min} = 937.993$ MeV. The analysis incorporates data from binary pulsar systems PSR J0348+0432, and PSR J1614--2230.
    }
\end{figure}

\section{Quasi-equilibrium Mass-Loss in Binary Pulsars}
\label{app:formalism}
In this section, we provide a concise overview of the quasi-equilibrium BNV formalism developed in Ref.~\cite{Berryman:2022zic}. 

The structure of a neutron star can be approximated by a static and spherically symmetric metric ($g_{\mu\nu}$) with a line element given by~\cite{tolman1987relativity}
\begin{equation}
\label{eq:g:def}
    d\tau^2 = g_{\mu\nu} dx^{\mu} dx^{\nu} = e^{2\nu(r)} \, dt^2 - e^{2\lambda(r)}\, dr^2 - r^2\, d\theta^2 - r^2\, \sin^2\theta\, d\phi^2,
\end{equation}
in which $\nu(r), \lambda(r)$ are solutions to the Einstein field equations~\cite{Einstein:1916vd}, $G^{\mu\nu} = -8 \pi G T^{\mu\nu}$, in which $G^{\mu\nu}$ is Einstein's tensor, $G$ is the gravitational constant, and $T^{\mu\nu}$ is the stress-energy tensor. We also assume that the medium in neutron star can be described by a perfect fluid with 
\begin{equation}
    \label{eq:macro_bnv:assum:tensor:perfect}
    T_{0}^{\,0} = {\cal E}, \quad T_{i}^{\,i} = -P \qquad (i = 1,2,3),
\end{equation}
as the only nonzero components of the stress-energy tensor in which $P$ and ${\cal E}$ are the local pressure and energy density of the fluid respectively which in general depend on the local baryon number density ($n$) and temperature ($T$) via the EoS.

The baryon decay rate (per baryon) in a small volume ($V$) in the nuclear matter (nm) rest frame ($\Gamma_{\rm nm}$) is defined by $d(n V)/d\tau = - \Gamma_{\rm nm} \, n \, V$, in which $\tau$ is the fluid's proper time, and $n$ is the proper baryon number density. The total rate of baryon loss in a neutron star is given by
\begin{equation}
    \frac{\dot B}{4\pi} = -\int e^{\nu(r)}\, \left[1 - \frac{2M(r)}{r}\right]^{-\frac{1}{2}}\, \Gamma_{\rm nm}(r)\, n(r)\, r^2\, dr, \label{eq:B:rate}
\end{equation}
where $B$ is the total baryon number of the neutron star. We have used $\sqrt{-g} = \exp(\nu(r)+\lambda(r))\, r^2\, \sin\theta$, with $\exp(2\lambda(r)) = (1 - 2M(r)/r)^{-1}$, and $M(r)$ is the total mass included within radius $r$:
\begin{equation}
\label{eq:m:def}
M(r') = 4\pi \int_0^{r'} {\cal E}(r) r^2 dr.
\end{equation}

If BNV is active in one or both of the components of a binary system, it can modify the rate of change in the orbital period via energy loss. Assuming that the BNV rate ($\Gamma_{\rm BNV}$) is much slower than the baryon-conserving responses of the neutron star, it will induce a quasi-equilibrium evolution of the star, i.e., along a path of equilibrium solutions. Furthermore, If no new particles, beyond those included in the Lagrangian that was used to derive the equation of state (EoS), are accumulated as a result of the BNV interactions, then the equilibrium solutions mentioned above are derived using the original (baryon-conserving) EoS. The BNV rate can then be constrained in a manner which is independent of the particle physics details of BNV, within this quasi-equilibrium BNV formalism~\cite{Berryman:2022zic}. 

Quasi-equilibrium BNV at a rate $\dot{B}$ contributes to an energy loss $\dot{M}^{\rm eff}$ from neutron stars as~\cite{Berryman:2022zic}:
\begin{equation}
\label{eq:observables:mdot:eff}
    \dot{M}^{\rm eff} = \left( \partial_{{\cal E}_c} M +  \left(\frac{\Omega^2}{2}\right) \partial_{{\cal E}_c} I \right) \left(\frac{\dot{B}}{\partial_{{\cal E}_c} B} \right),
\end{equation}
where $\partial_{{\cal E}_c}$ represents the partial derivative with respect to the central energy density of stars, $M$ is the mass, $I$ denotes the moment of inertia, and $\Omega$ is the angular spin frequency. This energy loss affects the binary pulsar orbital period~\cite{1963ApJ...138..471H,10.1093/mnras/85.1.2, 10.1093/mnras/85.9.912}:
 \begin{equation}
 \label{eq:observables:jeans:mloss}
    \dot{P}_b^{\dot{E}} = - 2 \left(\frac{\dot{M}_1^{\rm eff} + \dot{M}_2^{\rm eff}}{M_1 + M_2}\right)\, P_b,
 \end{equation}
 in which $1$ and $2$ represent binary components. The observed orbital period decay rate, \(\dot{P}_b^{\rm obs}\), combines gravitational radiation (\(\dot{P}_b^{\rm GR}\)), intrinsic energy-loss (\(\dot{P}_b^{\dot{E}}\)), and extrinsic factors (\(\dot{P}_b^{\rm ext}\)): 
\begin{equation}
\dot{P}_b^{\rm obs} = \dot{P}_b^{\rm GR} + \dot{P}_b^{\dot{E}} + \dot{P}_b^{\rm ext}.\label{eq:binary_period}
\end{equation}
Using the measured and estimated values for each of these contributions, we can determine the $2\sigma$ limit on $\dot{P}_b^{\dot{E}}$. This limit can then be translated into a constraint on the rate of baryon loss, $\dot{B}$, as detailed in Table~\ref{tab:psrbinary}, by applying Eqs.~\eqref{eq:observables:jeans:mloss} and \eqref{eq:observables:mdot:eff}.

\section{Escape Probabilities of $\psi$ Particles}
\label{app:escapee}

In this section we show that all of the heavy $\psi$ particles ($m_{\psi} \gtrsim 1\, {\rm GeV}$) produced in baryon decays ${\cal B} \to \psi + \gamma$ become gravitationally bound to the neutron star. Therefore, each of their subsequent scattering, as shown in Fig.~\ref{fig:charged_meson_scattering_feynman}, leads to one additional unit of baryon loss. We first express the escape condition at radius $r$, and in terms of $\psi$'s energy in the global neutron star frame $E_{\psi}^{\rm NS}(r)$ as
\begin{equation}
E_{\psi}^{\rm NS}(r) =  e^{+\nu(r)} E_{\psi}^{\rm nm}(r)    >  E_{\rm esc}^{\rm NS}(r) =  m_{\psi} e^{-\nu(r)}, \label{eq:escape:cond:energy}
\end{equation}
where the $\psi$'s energy in the local nm-frame, $E_{\psi}^{\rm nm}$, is given by
\begin{equation}
    E_{\psi}^{\rm nm} = \frac{E_{\cal B}}{2} \left[ 1 + \frac{m_{\psi}^2}{E_{\cal B}^2 - |\vec{p}_{\cal B}|^2} + \frac{|\vec{p}_{\cal B}|}{E_{\cal B}} \left(1 - \frac{m_{\psi}^2}{E_{\cal B}^2 - |\vec{p}_{\cal B}|^2}\right) \cos\theta\right],
\end{equation}
in which all the quantities on the right hand side are defined in the nm frame, and $\theta$ is the angle between $\vec{p}_{\cal B}$ and the direction of $\psi$ in the baryon's cm frame $\vec{p}_{\psi}^{\, \rm cm}$. We note that those $\psi$'s that are produced in the same direction as $\vec{p}_{\cal B}$, have a higher chance of escaping the star. Therefore, the escape condition in Eq.~\eqref{eq:escape:cond:energy} can be written in terms of $\cos \theta$ as 
\begin{equation}
    1 \geq \cos \theta \geq \frac{ 2 E^{\rm nm}_{\psi}/m^*_{\cal B}  - (x+\sigma) \left[1 + \mu_{\psi}^2 / (1 + \sigma^2 + 2 x \sigma)\right]}{ \sqrt{x^2 - 1} \left[ 1 -  \mu_{\psi}^2 / (1 + \sigma^2 + 2 x \sigma)\right] } \equiv \cos \theta_{\rm esc} ,
\end{equation}
in which we utilized the defintions in Eq.~\eqref{eq:medium:dim_less_def}. For any given baryon for which $\cos \theta_{\rm esc} \geq 1$, none of the $\psi$s can escape and if $\cos \theta_{\rm esc} \leq -1$, all of them would escape. Put more concretely the escape rate of $\psi$'s (per baryon) in the nm frame is given by
\begin{equation}
\begin{split}
    \Gamma_{\rm nm}^{\rm esc}= &\frac{\varepsilon_{{\cal B}\, \psi}^2\, g_{\cal B}^2\, e^2 \left(m_{\cal B}^*\right)^2}{128 \pi^3} \int_{1}^{x_F} dx \sqrt{x^2 - 1} \left[ \frac{1+ 2x\sigma + \sigma^2 - \mu^2}{\left(1 + 2 x \sigma + \sigma^2\right)^2}\right] \\
    &\times \frac{1}{2}\int_{\cos \theta_{\rm esc} }^{+1} d\cos\theta
    \bigg[ 
    \left(1 + 2 x \sigma + \sigma^2\right)\left(1 + x \sigma + 2 \mu\right) + \mu^2 \left(1 + x \sigma\right)\\
    &\qquad\qquad\qquad\qquad\quad -\sigma \sqrt{x^2-1} \left( 1 + 2 x \sigma + \sigma^2 -\mu^2 \right)\frac{\sin^2 \theta}{2} 
    \bigg].
\end{split}
\end{equation}

\begin{figure}[h!]
 \centering
    \includegraphics[width=0.55\linewidth]{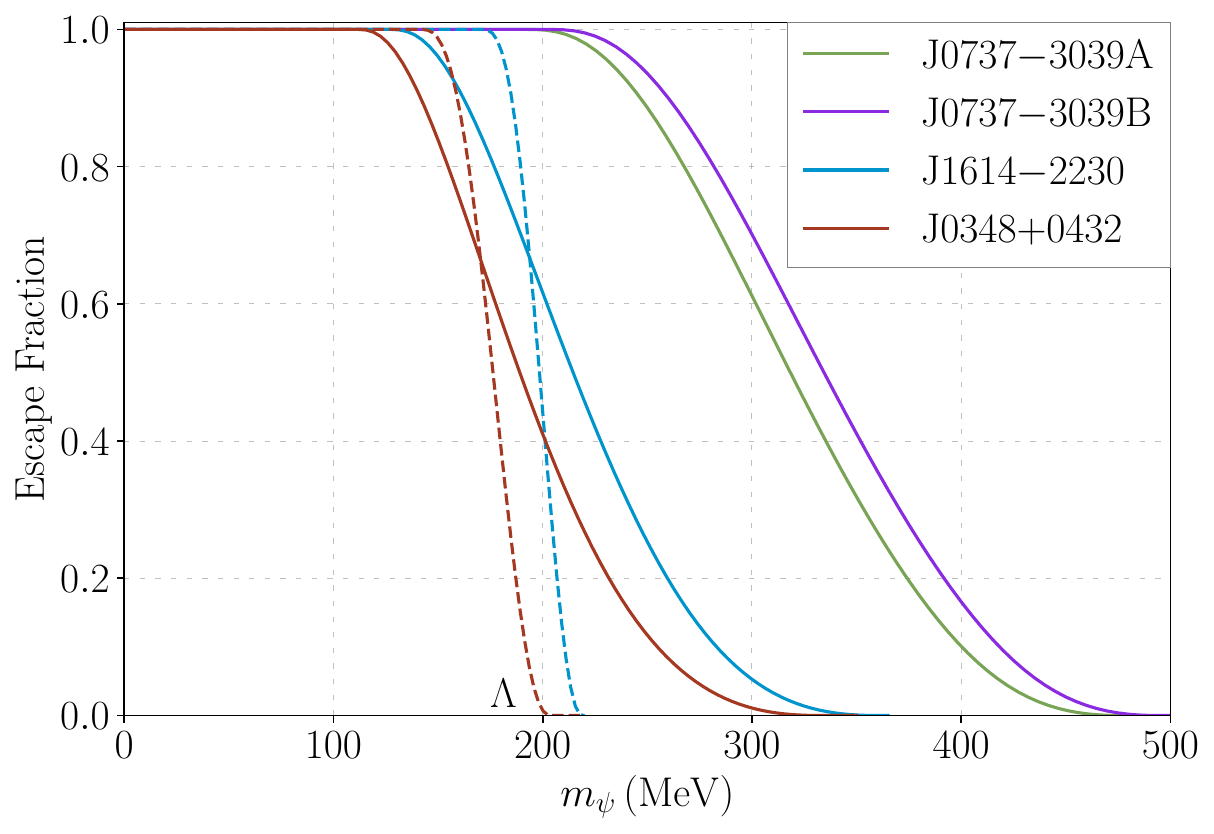}
    \caption{The fraction of $\psi$ particles, produced in $n \to \gamma \psi$ decays, that can escape pulsars is plotted using solid curves as a function of $m_{\psi}$. The dashed curves correspond to $\psi$ particles produced in $\Lambda \to \gamma \psi$ decays for the heavier pulsars J1614$-$2230 and J0348$+$0432.}
    \label{fig:escape_fraction}
\end{figure}

The total escape fraction for each pulsar can then be defined using Eq.~\eqref{eq:B:rate} as follows
\begin{equation}
    \text{Escape Fraction} \equiv \frac{\int e^{\nu(r)}\, \left[1 - \frac{2M(r)}{r}\right]^{-\frac{1}{2}}\, \Gamma_{\rm nm}^{\rm esc}\, n(r)\, r^2\, dr}{\int e^{\nu(r)}\, \left[1 - \frac{2M(r)}{r}\right]^{-\frac{1}{2}}\, \Gamma_{\rm nm}\, n(r)\, r^2\, dr}, \label{eq:escape_frac}
\end{equation}
in which $\Gamma_{\rm nm}$ is the total $\psi$ production rate in the nm frame. The result of Eq.~\eqref{eq:escape_frac} for different pulsars in plotted in Fig.~\ref{fig:escape_fraction}. We can see that for $m_{\psi} \gtrsim 500\, {\rm MeV}$ all of $\psi$s are bound to the pulsars we consider.

\end{document}